%% file: MaterialCharacterization_FEA_4DPrinting.tex
\documentclass[preprint,5p,times,twocolumn]{elsarticle}
\usepackage{amssymb}
\usepackage{textcomp}
\journal{Computer-Aided Design}
\usepackage{graphicx}
\usepackage{amsmath,amsfonts,txfonts}
\usepackage{epsfig}
\usepackage{graphicx}
\usepackage{amsmath}
\usepackage{amssymb}
\usepackage{subcaption}
\usepackage{xcolor}
\usepackage{todonotes}
\usepackage{lscape}
\graphicspath{{Fig/}}

\begin{document}
\begin{frontmatter}



\title{Material Characterization and Precise Finite Element Analysis of Fiber Reinforced Thermoplastic Composites for 4D Printing}

\author[label1]{Yuxuan Yu\corref{cor2}}
\author[label1,label2]{Haolin Liu\corref{cor2}}
\author[label1]{Kuanren Qian}
\author[label2]{Humphrey Yang}
\author[label2]{Matthew McGehee}
\author[label2]{Jianzhe Gu}
\author[label2]{Danli Luo}
\author[label2]{Lining Yao\corref{cor1}}
\author[label1]{Yongjie Jessica Zhang\corref{cor1}}

\address[label1]{Computational Bio-modeling Laboratory, Department of Mechanical Engineering, Carnegie Mellon University}
\address[label2]{Morphing Matter Lab, Human-Computer Interaction Institute, School of Computer Science, Carnegie Mellon University}

\cortext[cor1]{Corresponding authors \\\hspace*{1.6em} E-mail address: \{liningy, jessicaz\}@andrew.cmu.edu}

\cortext[cor2]{Contributed equally}

\begin{abstract} 
Four-dimensional (4D) printing, a new technology emerged from additive manufacturing (3D printing), is widely known for its capability of programming post-fabrication shape-changing into artifacts. Fused deposition modeling (FDM)-based 4D printing, in particular, uses thermoplastics to produce artifacts and requires computational analysis to assist the design processes of complex geometries. However, these artifacts are weak against structural loads, and the design quality can be limited by less accurate material models and numerical simulations. To address these issues, this paper propounds a composite structure design made of two materials - polylactic acid (PLA) and carbon fiber reinforced PLA (CFPLA) - to increase the structural strength of 4D printed artifacts and a workflow composed of several physical experiments and series of dynamic mechanical analysis (DMA) to characterize materials. We apply this workflow to 3D printed samples fabricated with different printed parameters to accurately characterize the materials and implement a sequential finite element analysis (FEA) to achieve accurate simulations. The accuracy of deformation induced by the triggering process is both computationally and experimentally verified with several creative design examples, and the 95\% confidence interval of the accuracy is $(0.972, 0.985)$. We believe the presented workflow is essential to the combination of geometry, material mechanism and design, and has various potential applications.
\end{abstract}


\end{frontmatter}

\section{Introduction}
\label{intro}
Starting from the last century, fused deposition modeling (FDM) with additive manufacturing (3D printing) has become a widespread approach to build new structures \cite{ngo2018additive,gao2015status}. Many types of thermoplastics, such as \textit{polyether ether ketone (PEEK)} and \textit{polyether ketone ketone (PEKK)}, are frequently used in 3D printing technology for some high-end applications~\cite{hu2019improved,singh2019investigations,roskies2017three}. By feeding thermoplastic filament through the heated printing nozzle, the melted material is squeezed out and falls onto a horizontal low-temperature platform; specific printing paths along with printing parameters are designed and programmed to control the movement of the printing nozzle to form the desired 3D structure. As the deposited material accumulates, the structure is built layer by layer. Based on the programmability of materials and printing paths, 3D printing changes the manufacturing process and provides an alternative approach for designers who prefer rapid structure modeling with customized materials and detailed distribution.

While the 3D printing industry has been introduced for both micron-length fibers and continuous fiber composites~\cite{tian2016interface}, printing self-assembled structures has not yet been fully adopted. The concept of self-assembly or the term \textit{4D printing} \cite{tibbits20144d}, which uses shape memory materials (SMM) \cite{ngo2018additive} and post-production actuation to combine 3D printing and time, is highlighted for its capability to fabricate adaptive structures \cite{tibbits20144d,ge2013active,kwok2015four} and could be traced back to the self-folding concept \cite{deng2015origami,an2014end}. 4D printing produces artifacts with smart materials which enable shape-changing \cite{leung2019challenges} in specific environment \cite{deng2015origami, davis2016self, deng2017accurately}. Traditional 3D printing is capable of creating various geometry by tweaking the geometry and printing parameters, while 4D printing allows users to control the final shape of an adaptive structure with programmable configuration.

Despite the possibility of designing programmable structures with 4D printing techniques, there exist several limitations. From the designer's perspective, there are two main limitations related to the 4D printing workflow: materials not being stiff enough and numerical simulations not being accurate enough. For materials usually used in 4D printing, thermoplastics is one of the most popular shape memory polymers (SMP), which is renowned for its high flexibility and material property transition under high temperature. However, many printable thermoplastics, including polylactic acid (PLA), have poor mechanical performance, which is not favorable in structural design. For numerical simulations, finite element analysis (FEA) is a commonly used technique to predict the deformation of the designed structure \cite{kwok2017gdfe}. However, without high simulation precision, FEA results cannot truly help designers predict the structure's final configuration, and designers must keep testing until the final desired shape is achieved, which is extremely expensive and time-consuming \cite{bakarich20154d, ge2013active}. To facilitate the 4D printing design process, both material and simulation aspects need upgrading: materials used in 4D printing need to be strengthened, and prediction from simulation needs to be more accurate to speed up the forward design process. In this paper, we provide solutions to both aforementioned issues.

Carbon-fiber PLA (CFPLA) is known for its high stiffness-to-weight ratio and is commonly used as a reinforcement material. PLA is soft and easy-to-deform when heated above the glass transition temperature ($T_{g}$), which enables PLA to be the most widely used thermoplastics in 3D printing. To maintain the merits of flexibility and easy-processing, meanwhile elevating its mechanical performance, we devise the use of CFPLA. With CFPLA, the designed structure can be strengthened along the direction of carbon fiber alignment; since the matrix material of CFPLA is still PLA, it can be easily 3D-printed with slightly higher extrusion temperature. In addition, based on 4D printing techniques, CFPLA can be easily combined with PLA to form bi-layer fiber-reinforced composite (FRC), which is relatively more difficult to adopt traditional fabrication approaches~\cite{tian2016interface}. By programming the printing path, FRC can be conveniently and effectively designed and fabricated layerwise, and the accuracy of FRC design can also be improved.

Precise material characterization is important to ensure the accuracy of FEA. A lot of research has been conducted to characterize polymeric material properties \cite{grijpma1994co, maharana2009melt, langer1998drug}, including 3D printing filament materials like PLA and CFPLA. Soares \textit{et al.} \cite{soares2008constitutive} used an incompressible, isotropic neo-Hookean hyperelastic material to describe the mechanical response of stents. The static and dynamic loading effect on degradation of PLA stent fibers was studied to further define a hyperelastic incompressible material \cite{hayman2014effect}. Khan \textit{et al.} \cite{khan2013phenomenological} combined compressible Ogden hyperelastic model and generalized the Maxwell model to obtain the linear viscoelastic behaviors of biodegradable polymers. A modified Eyring energy was utilized to define viscoplastic behavior of PLA \cite{sontjens2012time}. Eswaran \textit{et al.} \cite{eswaran2011material} introduced anisotropicity into the material modeling. A shape memory model was used to define thermal expansion coefficients for each printing layer \cite{bodaghi20194d}.

In this paper, both PLA and CFPLA are tested based on repeated dynamic mechanical analysis (DMA), and specific models are subsequently proposed to characterize their material properties. Another cause of low FEA fidelity is the influence of residual stress and body force on 4D-printed samples during the triggering process. To improve the precision of simulation, we devise a new sequential FEA, which takes the influence of residual stress and body force into consideration, and minimizes the necessity of test-printing intermediate designs with an accurate prediction of the final shape from simulation results. With a high accuracy of the final configuration prediction, we propose a forward design process - or workflow - to summarize the iterative process from the material characterization to the FEA, and the design finalization and fabrication can be achieved by going through this proposed workflow. The main contributions of this paper can be summarized as follows:
\begin{enumerate}
    \item A novel workflow is proposed for forward design, with accurate material property characterization and precise FEA simulation. This workflow supports robust and accurate fabrication of the designed object through an iterative optimization process and accurate control of the final configuration.

    \item The material properties of 3D printing polymers, including both PLA and CFPLA, are characterized in a precise way based on the DMA experiments. The characterization results are effectively incorporated into FEA with accurate mathematical models.

    \item A sequential FEA is developed to achieve accurate simulation results, considering both the residual stress releasing and the body force creeping. We simulate these two processes in a sequence to precisely derive the final deformation of the fabricated product.

\end{enumerate}

The layout of this paper is as follows. Section \ref{sec:overview} overviews the workflow. Section \ref{sec:struct_fabri} briefly introduces the geometric scope, structure design and fabricating parameter determination. Section \ref{sec:material} mainly discusses the physics behind material properties and precise material characterization of both PLA and CFPLA. Section \ref{sec:FEA} introduces simulation modeling and sequential FEA. Section \ref{sec:result} shows some simulations and design results derived from the proposed workflow. Section \ref{sec:conclusion} draws conclusions and points out potential improvement and future directions.

\begin{figure}[!htb]
\center{\includegraphics[width=\linewidth]
{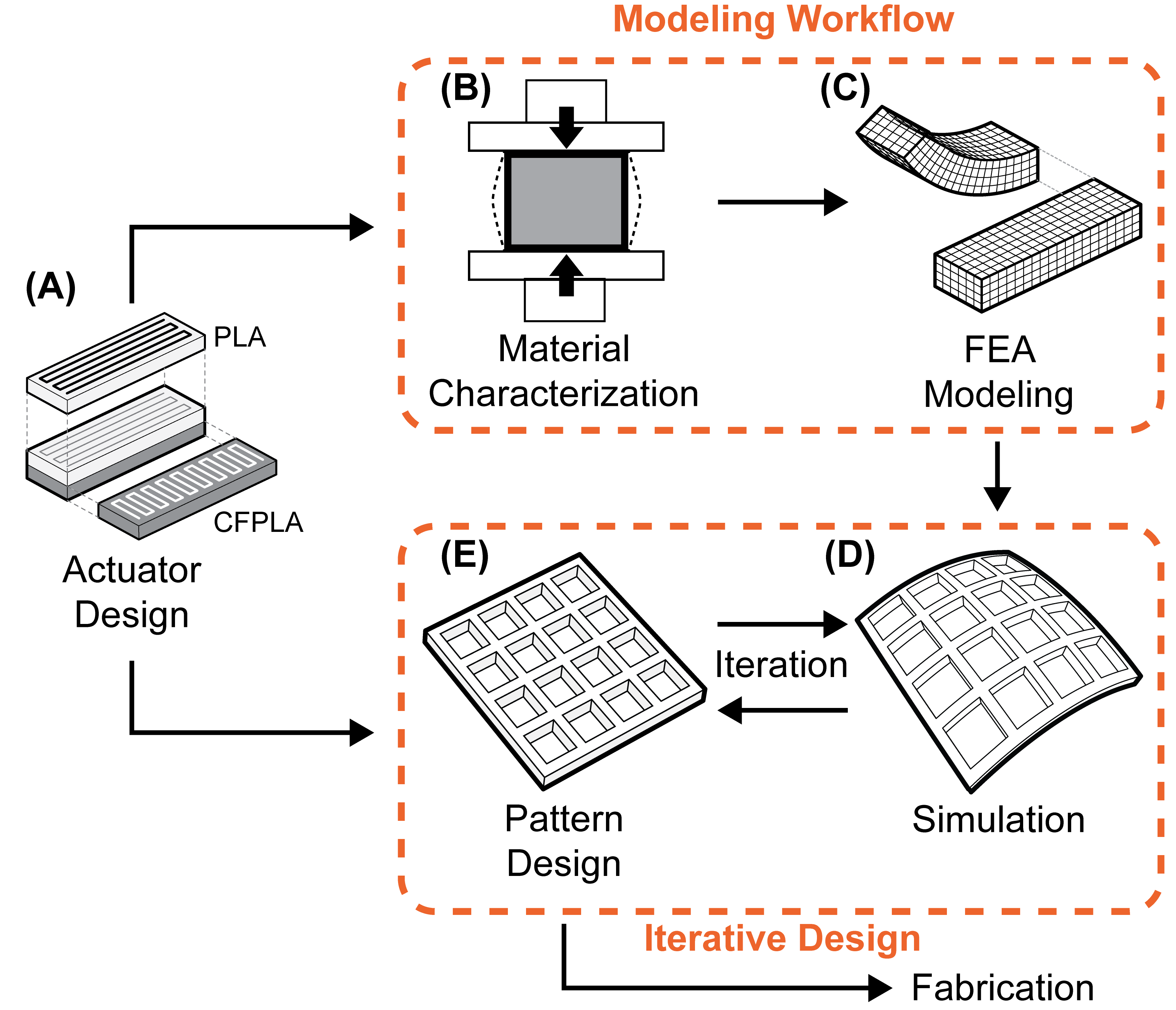}
\caption{\label{fig:Pipeline_Structure} The workflow of iterative 4D printing design. (A) shows the structure of a block with PLA as an actuator and CFPLA as constraint. (B) shows the process of material characterization based on DMA experiments. (C) refers to the FEA modeling on a bending unit. (D-E) show the iterative design process based on the simulation. This iterative process yields the finalized design, and leads to fabrication.}}
\end{figure}

\section{Overview of the workflow}
\label{sec:overview}
In this paper, a workflow is proposed to precisely simulate the self-morphing behavior of composite thermoplastic materials and structures. The proposed workflow consists of three major processes: material characterization, FEA modeling and simulation, and pattern design. As shown in Figure \ref{fig:Pipeline_Structure}, the workflow firstly characterizes the material property based on the DMA tests, and then creates FEA models and establishes material and boundary condition settings for Abaqus. In the end, the workflow simulates the final configuration to achieve the iterative forward design and fabricate the final product.

\textbf{Material characterization.} The material properties of PLA and CFPLA are characterized by using DMA experiments, which are conducted on strip samples under the temperature of 80\textdegree{C}, aiming to obtain the stress-strain curves and derive the material's hyperelasticity and viscoelasticity, respectively. The Mullins effect is taken into consideration to address material stress-softening~\cite{diani2009review}.

\textbf{FEA modeling and simulation.} Based on the aforementioned material characterization results, the material definition is determined in each FEA model. The residual stress values of each actuator and constraint are obtained by matching the triggering experimental results with their corresponding simulation results. Then, a sequential simulation is implemented to include both the residual stress releasing and the body force creeping. In this way, the simulation precision can be ensured and maintained at a high level.

\textbf{Pattern design}. With precise simulation results, designs can be created and refined iteratively. The simulation results from intermediate geometry can be obtained from the aforementioned FEA sequences, providing information on the difference between the current design and desired final configuration. An iterative process of forward design is then implemented from the initial geometry to the finalized design.

\section{Unit structure design and fabrication}
\label{sec:struct_fabri}
\begin{figure}[!htb]
\center{\includegraphics[width=\linewidth]
{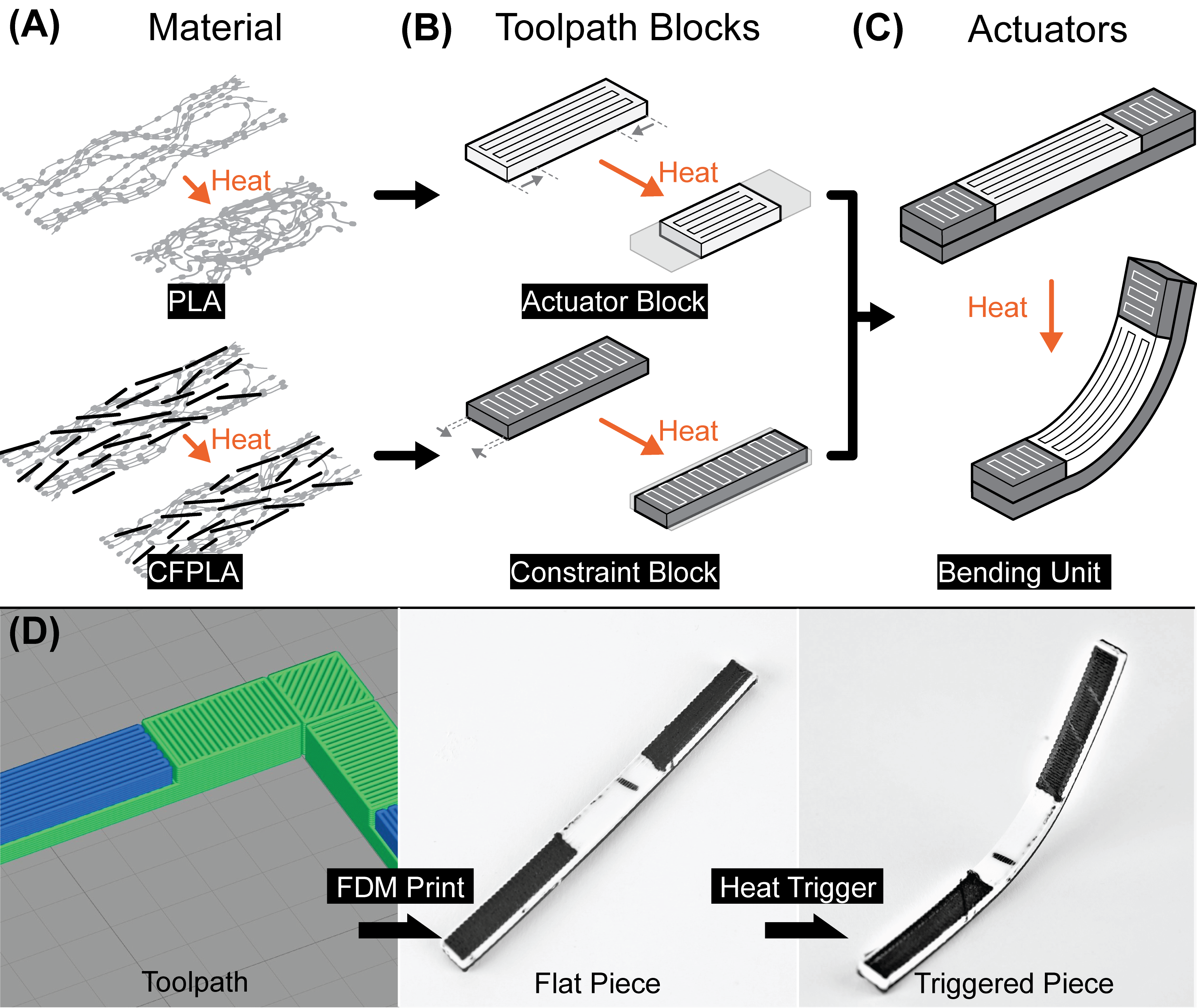}
\caption{\label{fig:PLA_Beam_Structure} Microscale view of PLA material and block structure with different functional components. (A) The polymer chain inside the 3D printed PLA structure. (B) Anisotropic blocks consisting of two layers with different printing path directions: the actuator blocks (white) and the constraint blocks (grey). (C) Bending unit with actuators (white) and constraints (grey). (D) The process from programmed printing path to fabricated flat piece and triggered bending unit.}}
\end{figure}

The theorem behind the deformation of 4D printing structures and materials is the residual stress releasing in response to external stimulation. 3D printed PLA materials release embedded residual stress when treated with a temperature higher than $T_{g}$;  see Figure \ref{fig:PLA_Beam_Structure}(A). The releasing of residual stress will result in the shrinkage of PLA along the printing path direction.

In this paper, all the designed products consist of bending units. As shown in Figure \ref{fig:PLA_Beam_Structure}(B-C), each unit is a cuboid with two different blocks - the top actuator block and the bottom constraint block, and they form the \textit{bi-layer unit} \cite{ionov2011soft,shim2012controlled,stoychev2013hierarchical}. Each block has a different printing path direction: the top block has a longitudinal printing path direction while the bottom block has a lateral printing path direction. The bending unit is the most fundamental structural unit in this work. When a bending unit is placed into an 80\textdegree{C} environment, the block with longitudinal printing paths will shrink more than the block with lateral printing paths, creating a difference in shrinkage and resulting in a bending behavior. The block with larger shrinkage is called \textit{an actuator block}, and the layer with less shrinkage is called \textit{a constraint block}.

Various structures and patterns are adopted in product design. In this study, the \textit{grid structure} is implemented for our design workflow, and its pattern is formed by bending units. The distribution of bending units with different actuators and constraints determines the deformation and configuration of the grid structure. Figure \ref{fig:PLA_Beam_Structure}(D) shows that each bending unit is assigned with a different actuator block and a constraint block, connected with pure constraint blocks as joints to form a grid. By controlling the length of the actuator block on each bending unit, different bending curvatures can be achieved and further combined to form a specified 3D shape.

\subsection{Three printing factors}
3D printing is the basic method we use for material processing and structure fabrication. The brand of 3D printer is \textit{Modix Big60}. To effectively control the deformation of the printed structure, filament materials and printing parameters are both tested. The distribution of actuator blocks and constraint blocks and the design of the structure should also be studied. In this paper, three printing factors with dominating effects on morphing characteristics are studied: printing path orientation, printing layer thickness, and filament material property.

\textbf{Printing path orientation} determines the shrinking direction of the thermoplastic material. During the FDM process, residual stress is embedded in PLA along the printing path as the filament is being extruded from the nozzle. As a result, the printed material shrinks along the specified printing path. Using this specific shrinking behavior of PLA material, different bending units can be created for design, as shown in Figure \ref{fig:PLA_Beam_Structure}.

\textbf{Printing layer thickness} also affects the shrinkage ratio. Thinner layer thickness requires more printing layers during the FDM process, which means more residual stress is embedded into PLA structures. As a result, higher shrinkage ratio can be achieved by thinner layer thickness. By combining blocks with different layer thicknesses into one bending unit, different shrinkage ratios are introduced to achieve various bending performance.

\textbf{Filament material property} significantly changes for different shrinkage ratios. The PLA and CFPLA filaments do not exhibit the same shrinkage ratio after the FDM process, because the CFPLA material is embedded with chopped up carbon fibers along the printing direction. The chopped-up fibers greatly improve the stiffness along the printing direction and significantly undermine the shrinking performance of the block.

\subsection{Determining optimal printing parameters}

To characterize material properties for different printing parameters through DMA experiments, we studied three types of thermoplastic composites. All experimental samples are straight bending units with the dimension of $75~mm \times 10~mm \times 4~mm$. Each sample consists of an actuator block at the top and a constraint block at the bottom. Actuator blocks are printed with a straight printing path that is along the sample's longitude direction, and constraint blocks are printed with a lateral printing path. After the printing process is completed, samples are fixed at one end and placed horizontally in a water bath at 80\textdegree{C} - which guarantees the homogeneous and stable heat transferring - to trigger the self-morphing process. In the experiments, three types of thermoplastic material designs were selected: PLA, CFPLA, and PLA-CFPLA combined bi-layer composite.

Experiments are also conducted to study the effects of the printing layer thickness on material bending behaviors. All experimental samples are printed with $0.6~mm$ nozzle,  $3,000~mm/min$ printing speed, but different layer thicknesses. $0.1~mm$, $0.2~mm$, $0.3~mm$, $0.4~mm$, and $0.5~mm$ are chosen as the printing layer thicknesses for three batches of sample blocks, categorized by different material types.

\begin{figure}[!htb]
\center{\includegraphics[width=\linewidth]
{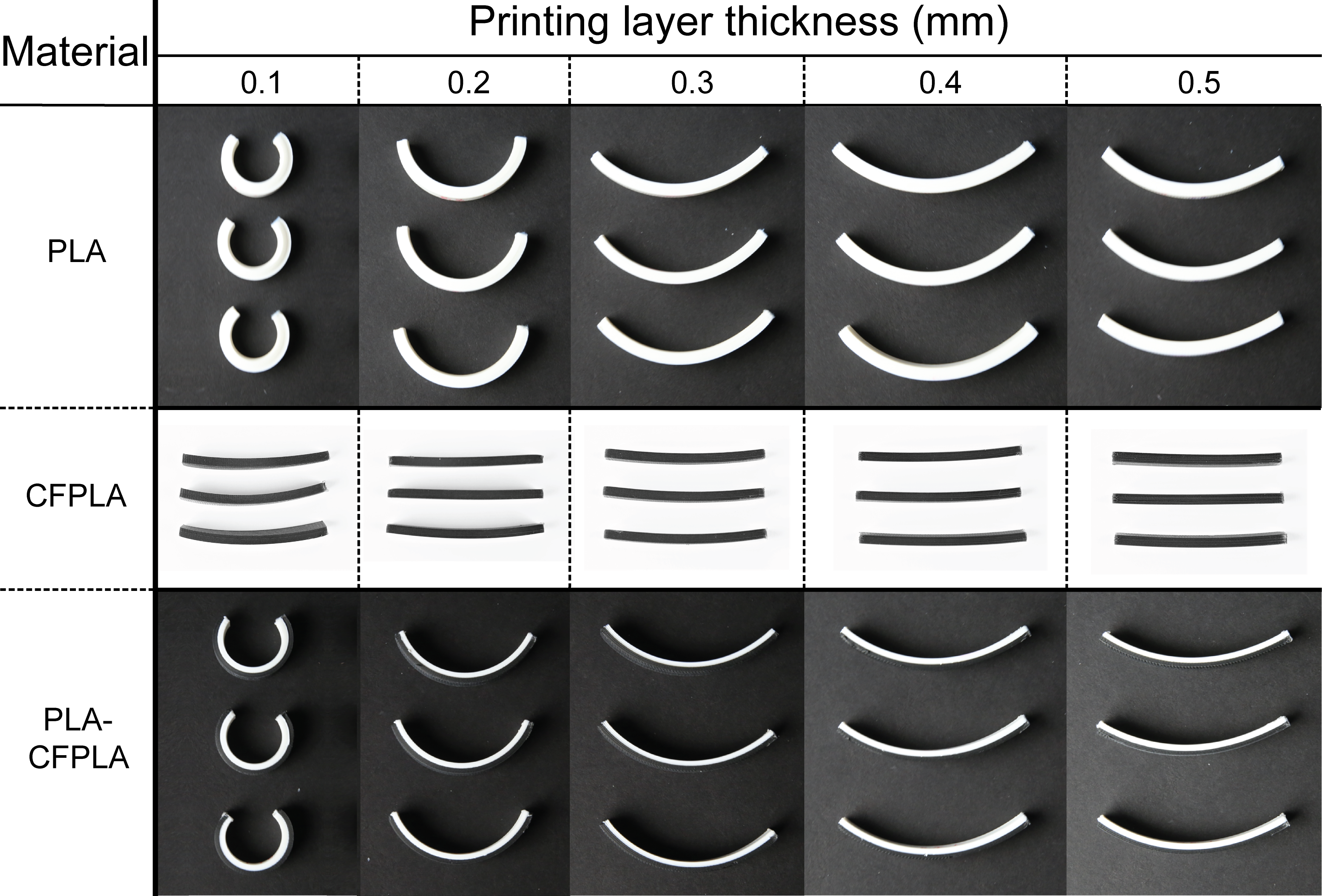}
\caption{\label{fig:All_Layer_Thickness} Experiment results of three types of thermoplastic materials: PLA, CFPLA and PLA-CFPLA composite bi-layer blocks, with varying layer thicknesses from $0.1~mm$ to $0.5~mm$. Results show that PLA bending units have the largest bending deformation while CFPLA bending units have the smallest bending deformation.}}
\end{figure}

\begin{figure*}[!ht]
  \centering{
  \includegraphics[width=\linewidth]{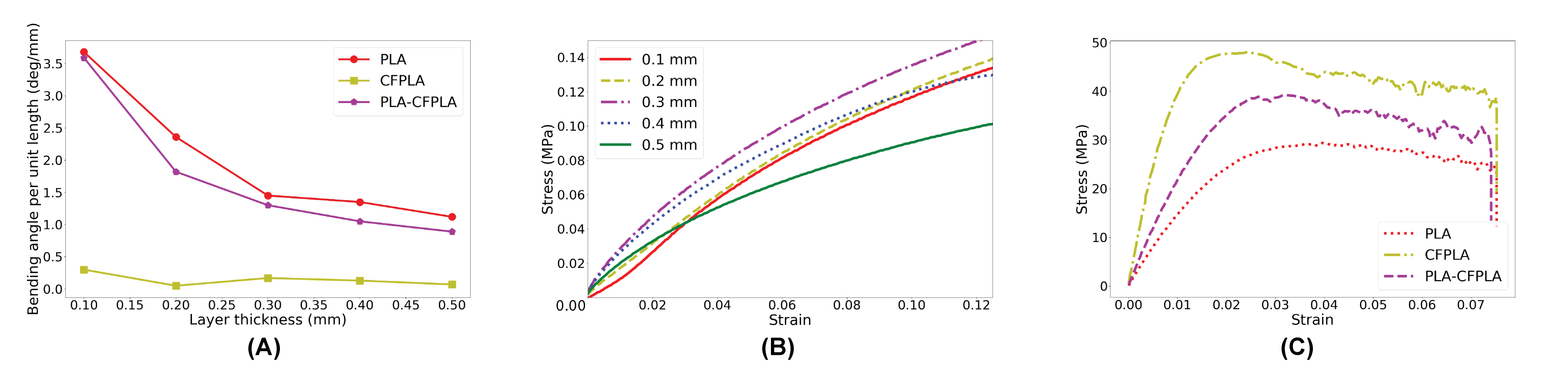}
  \caption{
    \label{fig:section3_long}
    The DMA experiment results for identifying the bending unit structure and the printing parameters. (A) Bending angle per unit length vs printing layer thickness for PLA, CFPLA and PLA-CFPLA composite. (B) Stress-strain curves of PLA-CFPLA composite for different printing layer thicknesses. Based on the experiments, $0.25~mm$ is chosen as the printing layer thickness. (C) Stress-strain curves with three different bending unit structures in the flexural test.}}
\end{figure*}

Figure \ref{fig:All_Layer_Thickness} presents the deformed shapes for experimental samples with varying layer thicknesses. We can observe that PLA samples with 0.1 mm printing layer thickness have the best bending performance among all PLA samples, while PLA samples with 0.5 mm printing layer thickness have inadequate bending performance. All CFPLA samples show minimal bending angles, indicating that the shrinkage ratio along the printing path direction of CFPLA is generally lower than that of PLA. Since CFPLA has limited bending performance, we choose the PLA-CFPLA composite bending units as the primary component in our design. This composite material is purposely designed to take advantage of the strength of CFPLA without sacrificing too much bending performance. Figure \ref{fig:section3_long}(A) shows that composite samples have similar shrinkage ratio and bending angle as PLA samples.

Figure \ref{fig:section3_long}(B) shows the stress-strain curve of PLA-CFPLA combined bending unit for five different printing layer thicknesses. Samples with printing layer thickness larger than $0.3~mm$ do not have a large enough bending angle; however, choosing a printing layer thickness $<0.2~mm$ results in extended printing time and unreliable stress-strain data. In this paper, we chose the \textit{Modix Big60} 3D printer for its versatility in printing sizes. However, the printer produces inconsistent results when dealing with layer thickness smaller than 0.2 mm. The \textit{Modix Big60} 3D printer, designed for large-size 3D printing tasks, has a printing volume of $61~ cm^{3}$ and implements a unified bed leveling technique to actively adjust for small dents across the aluminum printing bed during printing. The design tolerance for printing bed dents is 0.2 mm, according to the manufacturer's instructions. As a result, the automatic \textit{z}-height adjustment during printing leads to quality inconsistency if the printing layer thickness is smaller than 0.2 mm. Therefore, we choose the PLA-CFPLA combined bending units with $0.25~mm$ layer thickness for our workflow implementation.

\subsection{Flexural test}
Concerning the flexural test, three-point bending and four-point bending are conducted on bending units. The dimension of each unit sample for three-point bending test is $60~mm \times 4~mm \times 1.6~mm$, and the effective length of the supporting span is $40~mm$. The dimension of each unit sample for four-point bending test is $100~mm \times 8~mm \times 8~mm$, and the loading span and the supporting span are $23.56~mm$ and $70.68~mm$, respectively. Testing samples of PLA, CFPLA, and PLA-CFPLA composite are designed and tested with residual stress pre-released under room temperature to observe the corresponding flexural modulus and strength. The experimental setup of the three-point bending test is shown in Figure \ref{fig:Test-process}(C). Figure \ref{fig:section3_long}(C) shows the experimental result of the three-point bending in terms of flexural stiffness for three types of samples, and the corresponding four-point bending experimental result is listed in the supplementary material (Figure S3). From both experimental results, we can observe that although the CFPLA sample has the highest flexural stiffness, it is so brittle that it may break within relatively small effective deflection range; the PLA sample is capable of withstanding large deflection, but it has the lowest flexural stiffness among all three types of samples. By combining CFPLA and PLA into the bi-layer structure, the PLA-CFPLA composite bending unit can not only exhibits high flexural stiffness, but also withstand large deformation at the same time. Its strengthened mechanical properties are capable of expanding the application space of the designed structure under different loading cases.

\section{Material characterization}
\label{sec:material}
 Material characterization is important for precise simulation. Both PLA and CFPLA are characterized. The physics of PLA and CFPLA include their microstructure, anisotropic behavior and nonlinear and time-dependent material behavior. Based on the material physics, specific models are identified to facilitate FEA simulations.

\subsection{Material physics}
As a thermoplastic polymer, PLA is composed of numerous polymer chains \cite{jones2005engineering}. These chains are often entangled together, forming a micro-scale network. During the printing process, as the temperature arises above the PLA's melting point, slippage at the network's entanglement links may occur, introducing viscoelastic properties to the printed material. In the case of 4D printing, when we reheat the material over its glass transition temperature, these viscoelastic properties will play a dominant role in the material's deformation behaviors. Therefore in addition to the commonly observed plastic behaviors, the viscoelasticity of PLA must be considered in establishing an accurate material model.

\begin{figure}[!htb]
    \centering{
    \includegraphics[width=\linewidth]{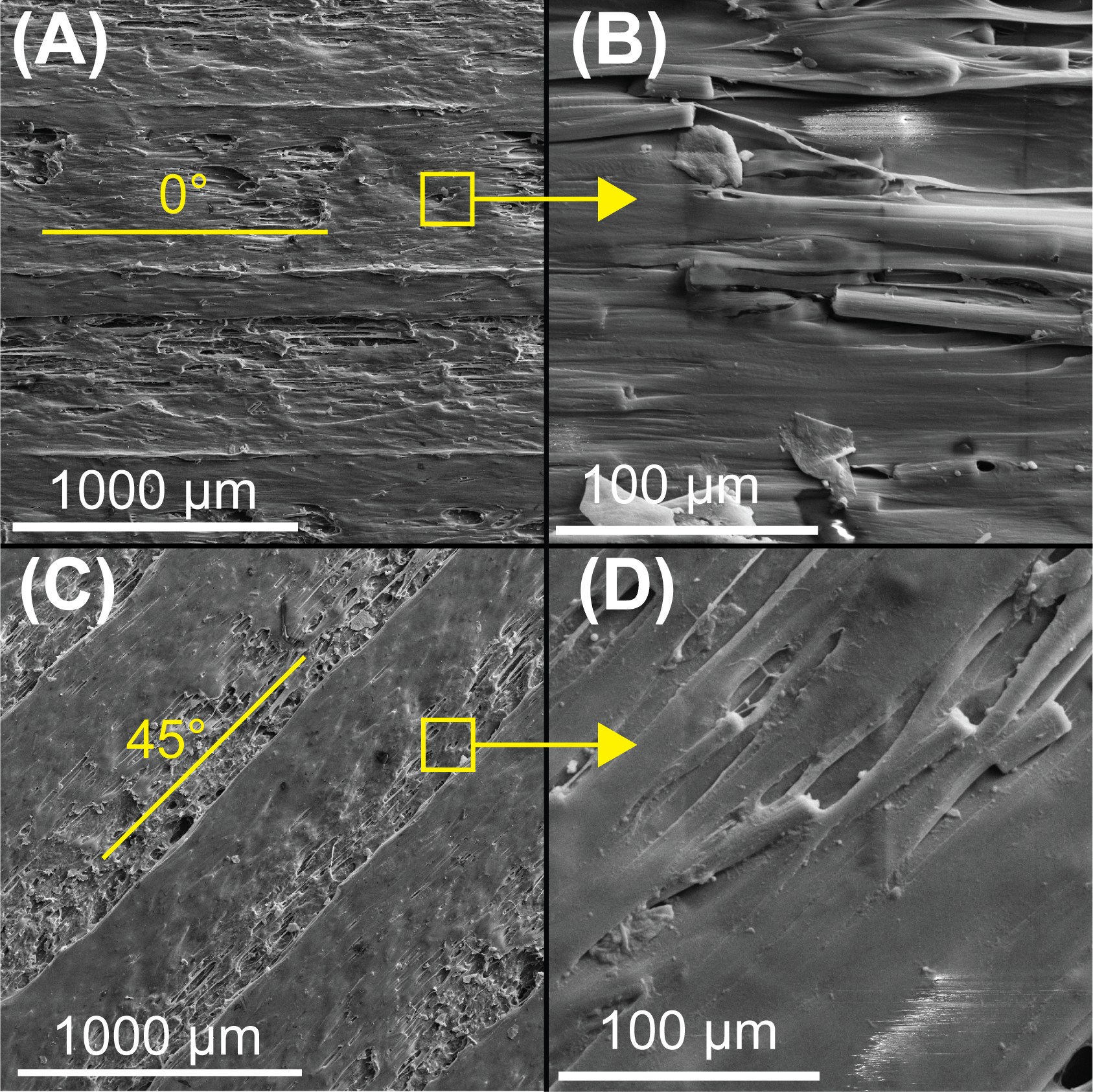}
    \caption{The SEM images of CFPLA. (A-B) show zoom-out and zoom-in pictures of the sample with the printing direction of 0\textdegree{}. (C-D) show zoom-out and zoom-in pictures of the sample with the printing direction of 45\textdegree{}.}
    \label{fig:SEM}}
\end{figure}

The polymer chain networks of PLA usually have random orientations, and the material can be considered to be isotropic. Yet, when the filament material is extruded from the nozzle during the printing process, the polymer chains are stretched along the printing direction, causing the material to become anisotropic. CFPLA also undergoes the same process during 3D printing, and fibers within the printed blocks are aligned, increasing the material's stiffness along the printing direction. Figure \ref{fig:PLA_Beam_Structure}(A) illustrates the detailed distribution and anisotropic material property of both PLA and CFPLA, and Figure \ref{fig:SEM} further evidences the fiber-alignment and anisotropy of printed materials with scanned electron microscopy (SEM) images.

\subsection{Material properties}
\label{sec:material_pro}
Based on the material physics, PLA is expected to show both elastic and viscous behaviors. We adopted a hyperelastic consitituive model to describe the hyperelastic behaviors, and collect data by performing the uniaxial tensile test on both PLA and CFPLA with the \textit{RSA-G2} equipment. The PLA material is purchased from \textit{Polymax}; the CFPLA material is purchased from \textit{Proto-Pasta}, with the average length of carbon fibers less than 150 microns. The following are the detailed testing procedures for each property of interest.

\begin{figure}[htb]
  \centering{
  \includegraphics[width=\linewidth]{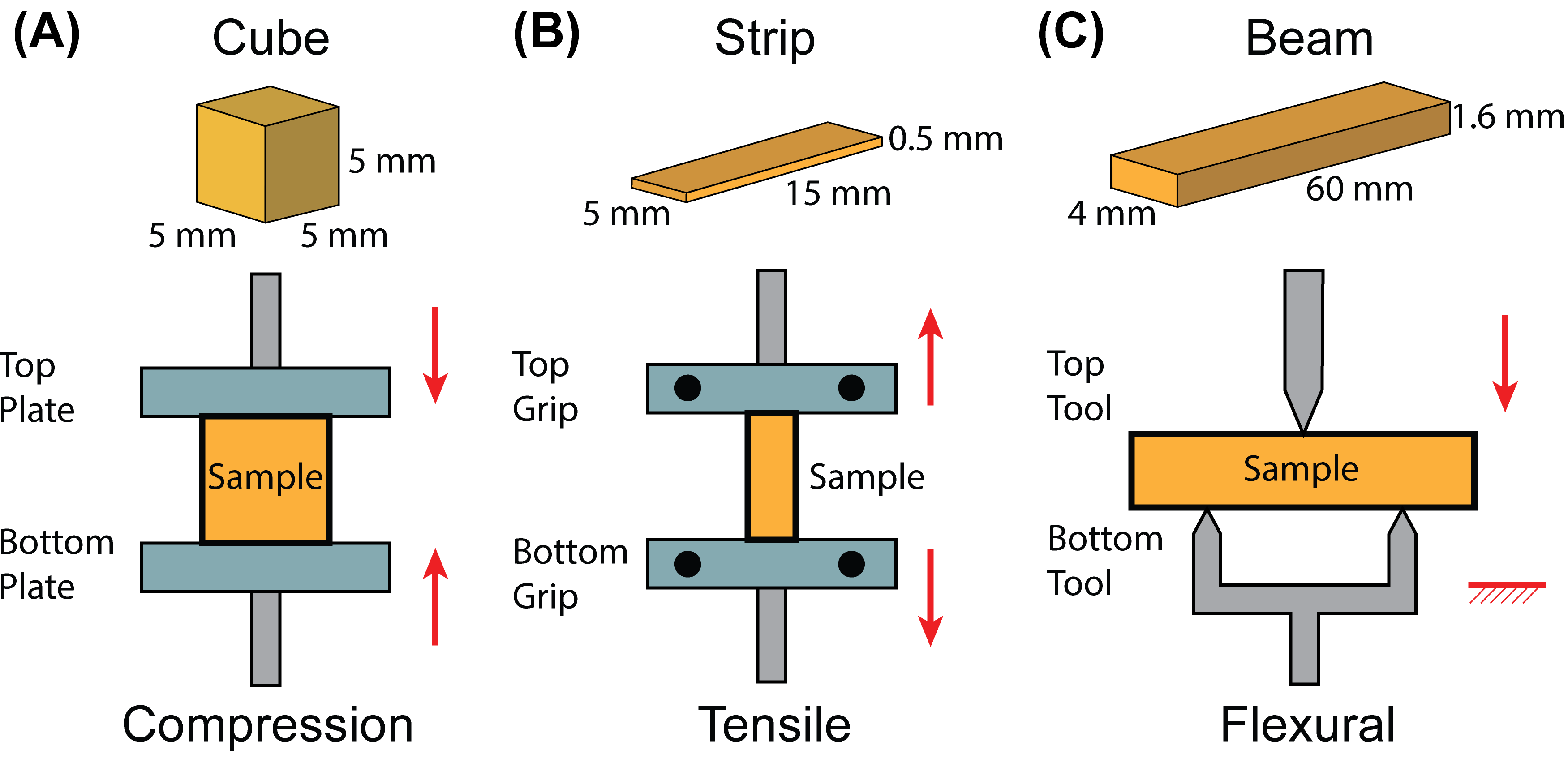}
  \caption{DMA experiments setup. (A) refers to the DMA compression test with a cubic sample. (B) refers to the tensile test with a strip like sample. (C) refers to the three-point bending test with a bi-layer block sample.}
    \label{fig:Test-process}}
\end{figure}

\textbf{Thermal expansion and Poisson's ratio}. Since the material and the structure are triggered in a high temperature, the material's thermal expansion rate ($\alpha_{t}$) and the Poisson's ratio ($\mu$) are considered in our workflow. These two parameters can be obtained by measuring the dimensional change of a 3D printed cubic sample before and after the uniaxial compression test. The cubic sample of size $5~mm \times 5~mm \times 5~mm$ is fabricated and subjected to a compressing load perpendicular to the top and bottom surfaces (Figure \ref{fig:Test-process}(A)). The obtained $\alpha_{t}$ and $\mu$ are shown as Table \ref{tab:compression}. Each result is with an error bound calculated from its $95\%$ confidence interval of measurement. 

\begin{table}[!ht]
    \centering{
    \caption{$\alpha_{t}$ and $\mu$ of PLA and CFPLA at 80\textdegree{C}}
    \vspace{-2mm}
    \begin{tabular}{c|c|c}
        \hline
        Material &  $\alpha_{t}$  (1/\textdegree{C}) & $\mu$ \\
        \hline
        PLA & $(9.17 \pm 1.54) \times 10^{-4}$ & $ 0.419 \pm 0.021 $ \\
        CFPLA & $(9.97 \pm 5.92) \times 10^{-5}$ & $ 0.359 \pm 0.015 $ \\
        \hline
    \end{tabular}
    \label{tab:compression}}
\end{table}

\begin{figure*}[!ht]
  \centering{
  \includegraphics[width=\linewidth]{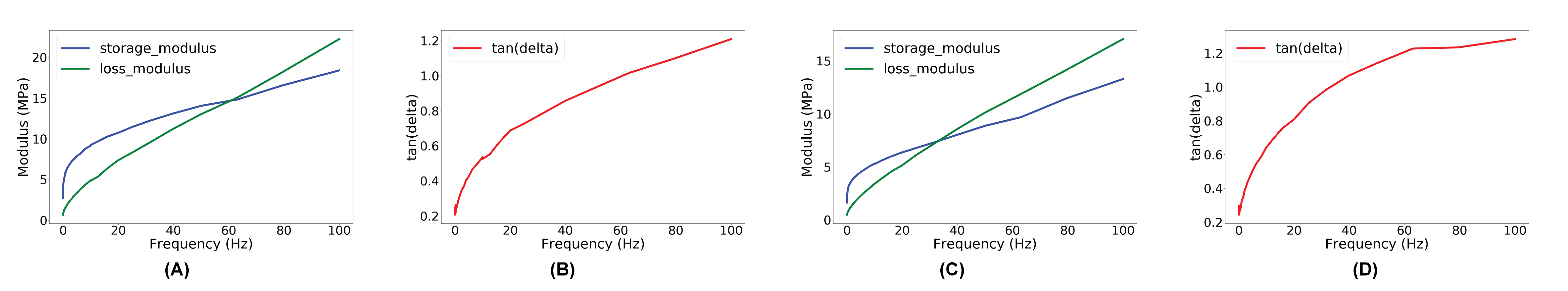}
  \caption{
    \label{fig:section4_long}
    Experimental results of the modulus-frequency plot and the damping curve for PLA actuator (A-B) and constraint (C-D).
    }}
\end{figure*}

\textbf{Hyperelasticity}. The hyperelasticity data of the materials are obtained by performing uniaxial tensile tests on a strip sample of size $5~mm \times 15~mm \times 0.5~mm$ (Figure \ref{fig:Test-process}(B)). Each sample is heated to release its residual stress under 80\textdegree{C}. Since PLA can reach the rubbery state above $T_{g}$, stress-softening or the Mullins effect are taken into consideration. To capture this effect, four cycles of force loading and unloading process are applied to the samples with a speed of $0.005~mm/s$ and a $23\%$ tensile strain limit. For CFPLA, due to its higher stiffness and smaller scale of deformation, the viscoelastic and plastic behaviors are insignificant during the triggering and deforming process. Therefore, we only consider the hyperelasticity for CFPLA by conducting the tensile test with a speed of $0.002~mm/s$ and a $3\%$ tensile strain limit, while the Mullins effect and plasticity are not considered in the CFPLA material setup.

\textbf{Plasticity}. The plastic behavior of PLA is observed after cycles of force loading period. This behavior is described by identifying the anchor point,  the intersection of the stress-strain curves and x-axis.

\textbf{Viscoelasticity}. We use the viscoelastic constitutive model to characterize the viscoelasticity of PLA. In the experimental setup (Figure \ref{fig:Test-process}(B)), the equipment pre-strains the samples ($0.0095$) and applies oscillating loads ($0.01~Hz$ to $100~Hz$) to record the storage $E'$ and the loss $E''$ modulus in the frequency domain to describe the frequency-dependent behaviors. Figure \ref{fig:section4_long} shows the modulus-frequency plot and the damping curve obtained for PLA actuator and constraint. The viscoelasticity experiment of CFPLA samples is also conducted, and the experimental result shows that the loss modulus to storage modulus ratio - which is also written as $tan\delta$ - of CFPLA is always less than 1, which indicates that the viscoelastic behavior of CFPLA is not dominant. Detailed evidence based on the experimental result is listed in the supplementary material (Table S3, Figures S1 and S2). 

\section{FEA modeling and sequential simulations}
\label{sec:FEA}
To successfully simulate deformation of designed products, FEA processes with different boundary conditions and initial conditions need to be implemented in order. In this paper, we propose two kinds of sequential simulations: the sequential residual stress estimation and the sequential deformation simulation.

\subsection{Sequential residual stress estimation}
\label{simulation_sequence}
Since the deformation of a bending unit results from its embedded residual stress, identifying the initial residual stress is extremely important. Numerically identifying residual stress can be convenient and programmable by implementing the shooting method \cite{osborne1969shooting}, which is a numerical method to obtain the initial residual stress and convert the boundary value problem to an initial value problem. Based on the shooting method, the process of residual stress estimation can be summarized in a sequence: (1) FEA modeling and material definition; (2) DMA experiment verification; and (3) residual stress and material property identification.

\subsubsection{FEA modeling and material definition}
\begin{figure}
    \centering{
    \includegraphics[width=\linewidth]{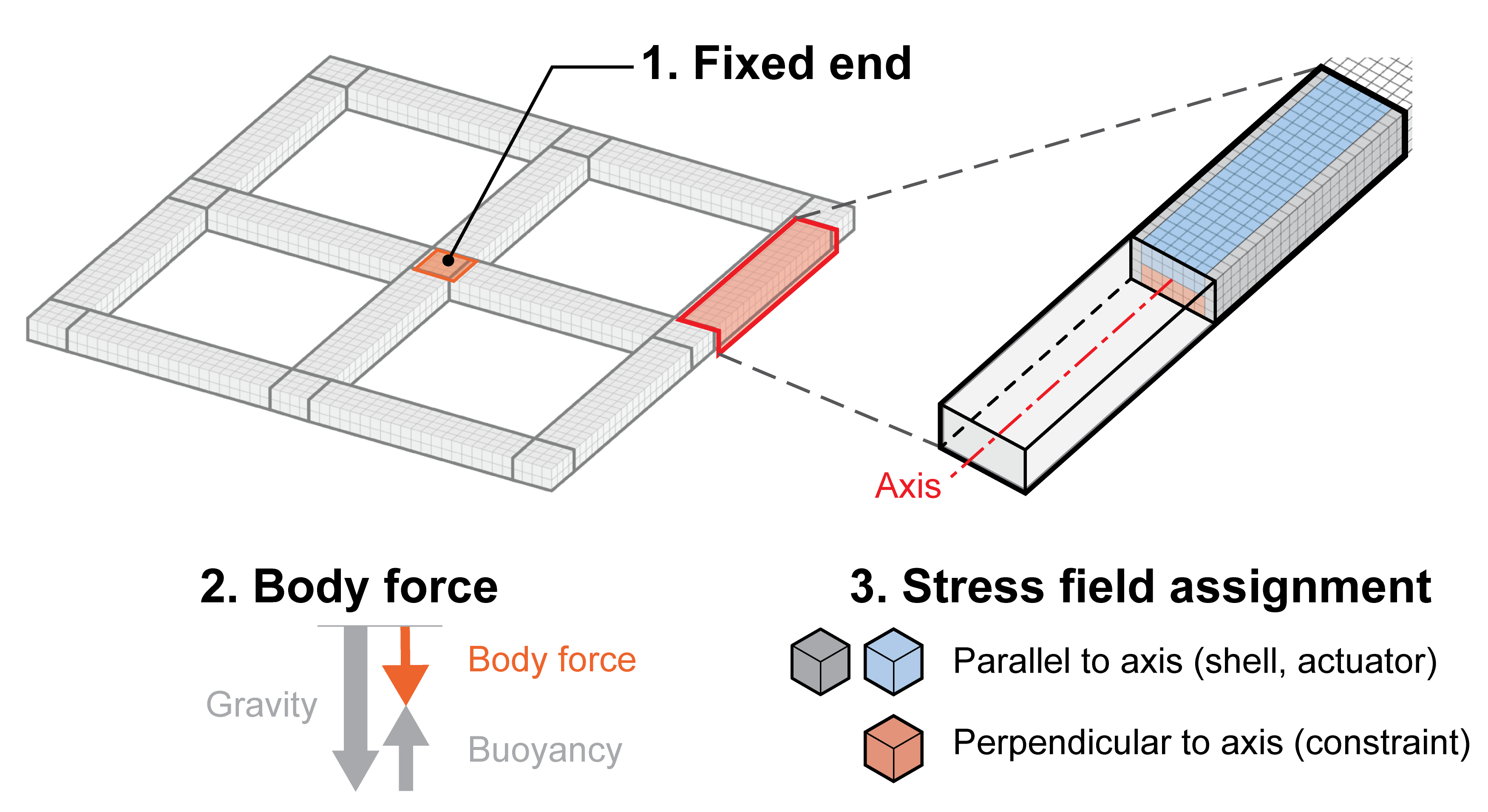}
    \caption{Model discretization and boundary condition settings of a simple grid structure.}
    \label{fig:model_disc}}
\end{figure}

FEA modeling includes meshing, boundary condition setting and material definition. As discussed in Section \ref{sec:struct_fabri}, the block and grid structure are the main structures studied in this paper, thus the geometry of all designed products can be easily discretized into hexahedral elements for FEA simulations in Abaqus. The boundary condition settings are consistent with the triggering experiment. Since the bending unit structure can be divided into actuator and constraint components, they are assigned with different material properties. The boundary condition settings are shown in Figure \ref{fig:model_disc}, including the fixed region, the body force and the division of actuator and constraint components. For the material property settings, we consider hyperelasticity, plasticity and viscoelasticity based on the raw data obtained from the aforementioned DMA experiments. 

\begin{figure}[htb]
  \centering{
  \includegraphics[width=\linewidth]{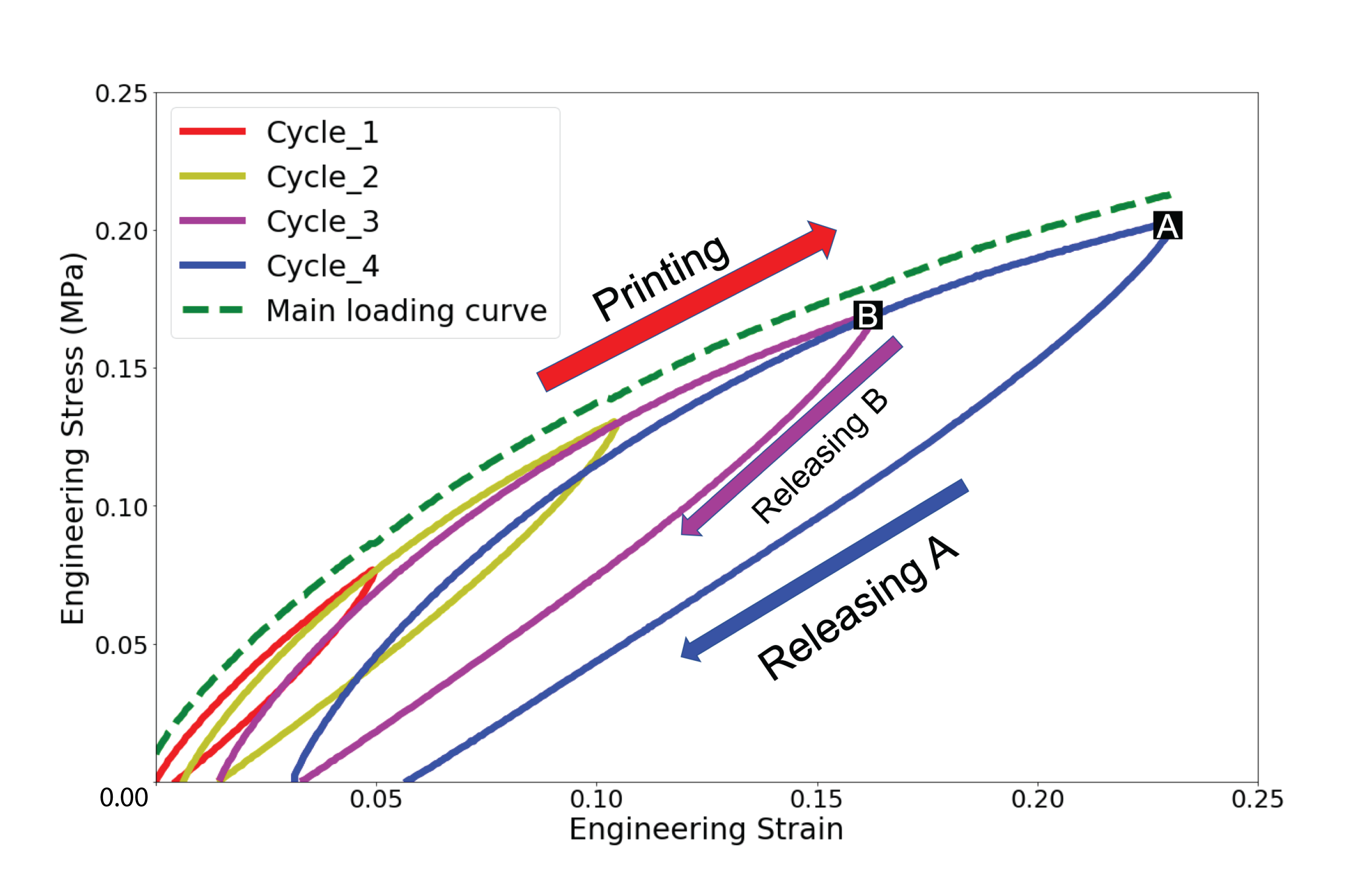}
  \vspace{-2mm}
  \caption{
    \label{fig:PLA_tensile_stress}
    The loading-unloading tensile test results for PLA. The stress-strain curves from four cycles of force loading and unloading clearly show the material's Mullins effect. The main loading curve can be obtained from the stress-strain curves marked as the green dashed line, which refers to the printing or residual stress embedding process. Different loading stresses (marked as A and B) yield different unloading stress-strain curves, which refer to the triggering or residual stress releasing process.
    }}
\end{figure}

In this paper, the material's \textit{hyperelasticity definition} is assumed to be isotropic. To remove noise in the raw stress-strain data from experiments, a penalty spline based smoothing method \cite{eilers1996flexible,yu2019anatomically} was implemented. The processed data for PLA is shown in Figure \ref{fig:PLA_tensile_stress}. The four cycles of loading (printing) and unloading (releasing) processes are shown, and the main loading curve (the green dashed line) can be derived from the raw data. The material's hyperelasticity is merely determined by the unloading curve, which is also relevant to the initial residual stress value. Different initial residual stresses may lead to different hyperelasticity definitions.

By using the aforementioned anchor points data, the material's \textit{plasticity definition} is also assumed to be isotropic. The unrecoverable deformation of the corresponding material can be obtained from the anchor points, and the Mises plasticity with an associated flow rule was implemented to model the material's plastic deformation \cite{lubliner2008plasticity}. In Abaqus,  the hyperelasticity and plasticity material properties are further combined. When only one set of data is available, the Marlow model is recommended \cite{systemes2019abaqus}. The Marlow strain energy potential \cite{marlow2003general} is chosen to fit the main loading curve, and the unloading curve is described by the Marlow model plus the Ogden-Roxburgh model as the damage \cite{ogden1999pseudo} and plasticity terms \cite{govindarajan2008simulation}. From Figure \ref{fig:PLA_tensile_stress}, we can observe that different stress values on the main loading curve correspond to different unloading curves, which also verifies that the material property during the stress-releasing process is highly dependent on the initial residual stress value.

The \textit{viscoelasticity definition} is obvious especially for PLA and is characterized by the storage modulus, the loss modulus, and the $tan\delta$ derived from the DMA experiments (see Section~\ref{sec:material}). Based on the explanation in Section \ref{sec:material_pro}, the viscoelasticity is not the dominant mechanical property for CFPLA material, thus we decide not to include the viscoelastic behavior into CFPLA's material definition.  

\vspace{2mm}
\textbf{Discussion 5.1.}  \textit{The Marlow model assumes that the strain energy potential is a function of the first strain invariant and is independent of the second strain invariant. Eliminating the second strain invariant from the strain energy potential has several benefits, as discussed in~\cite{yeoh1993some}. In our case, by using this model, we can reduce the number of experiments (the uniaxial tension test) required to describe the material behavior accurately. We can, therefore, avoid using complex experimental data from biaxial tension and planar tension for parameters calibration.}

\subsubsection{DMA experiment verification}
\begin{figure}[htb]
  \centering{
  \includegraphics[width=\linewidth]{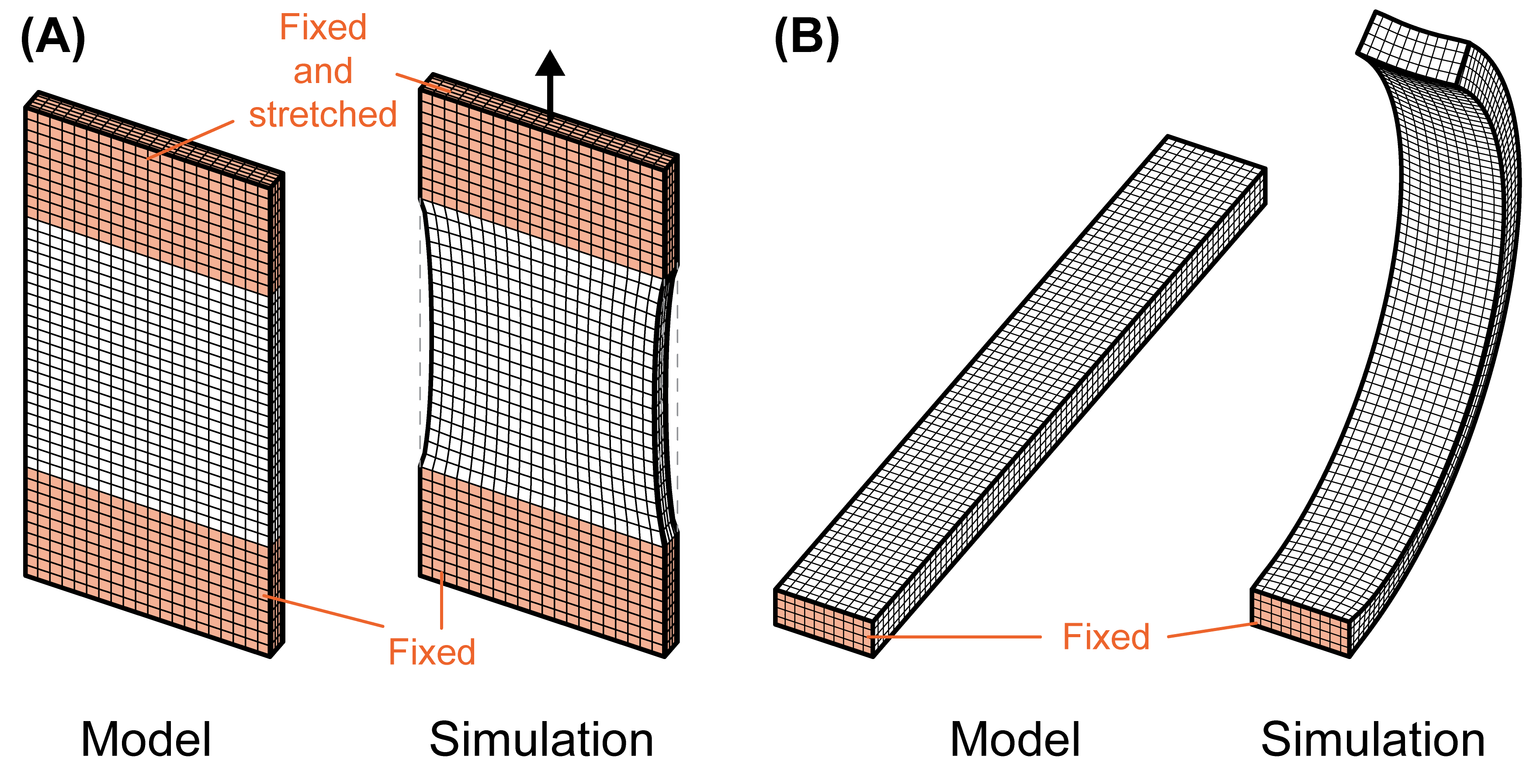}
  \caption{%
    \label{fig:loading_unloading}
    The tensile testing samples for DMA experiments (A) and the bending unit samples (B). The colored area shows the applied boundary conditions during experiments.
    }}
\end{figure}

\begin{figure}[htb]
  \centering{
  \includegraphics[width=\linewidth]{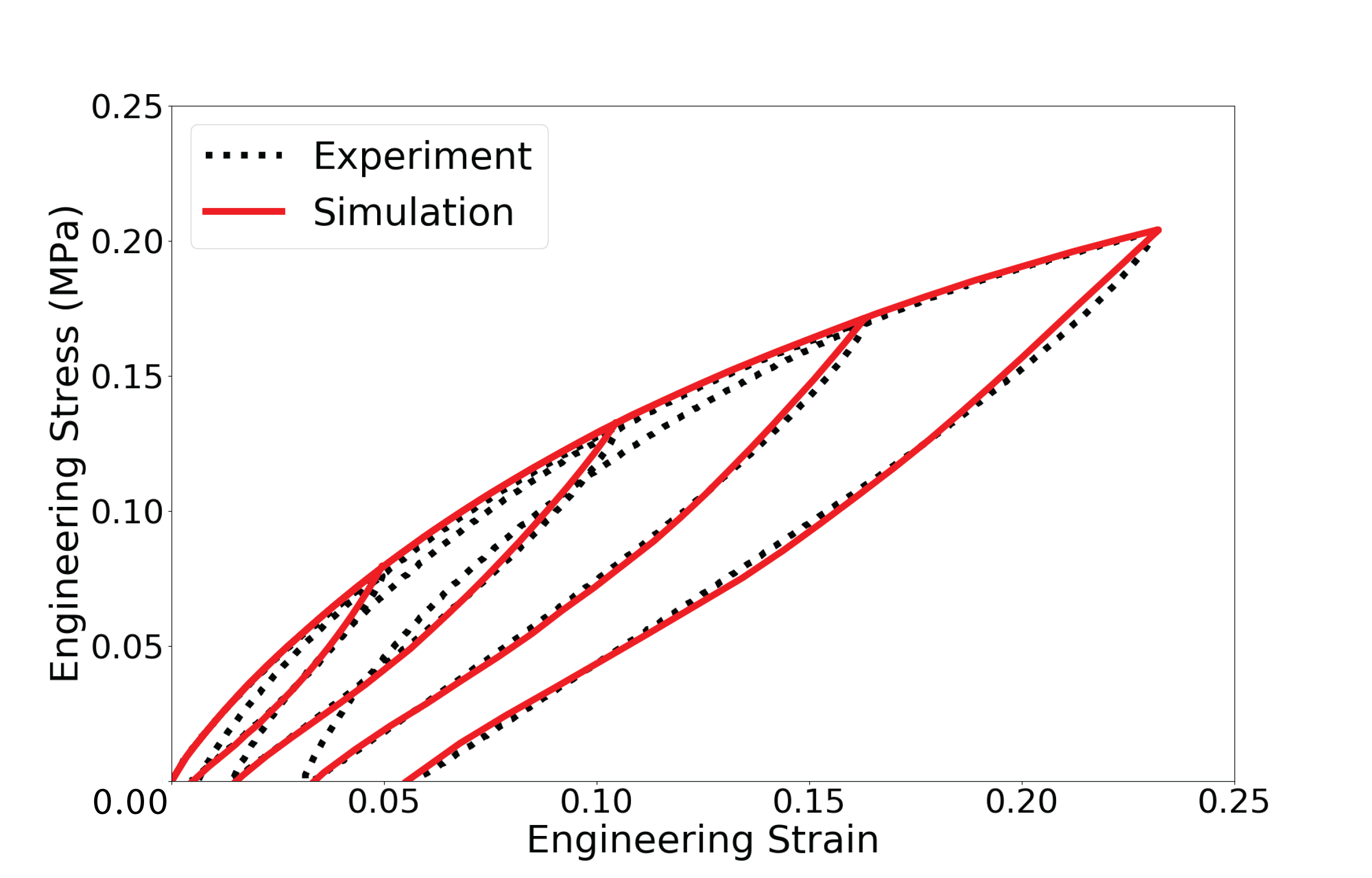}
  \vspace{-2mm}
  \caption{%
    \label{fig:PLA_tensile_stress_simulation}
    Comparison between the simulation results and the tensile test data under cyclic loading conditions.}}
\end{figure}

After quantifying the material properties, a simple simulation model of the strip sample in DMA experiments was created and tested to check how the simulation results match the DMA experimental results. The dimensions of the strip sample are shown in Figure \ref{fig:Test-process}(B), and the exact geometry is built for simulation. Fixing boundary conditions are set at the same position as in experiments, and the testing process can be simulated by applying a stretching load as shown in Figure \ref{fig:loading_unloading}(A). The simulated stress-strain results can be subsequently derived and compared with the experimental results as shown in Figure \ref{fig:PLA_tensile_stress_simulation}. We can observe that the simulation results match the experiment results pretty well. 

\subsubsection{Residual stress and material property identification}
To precisely estimate the initial residual stress and finalize the material property definition, an iterative process is programmed to implement the shooting method. To show different bending curvatures for different beams, the actuator ratio, which is the ratio of the actuator block length to the total length of the bending unit, is controlled from $0$ to $1$. The bending units with different actuator ratios should have different bending curvatures, which may help precisely identify the initial residual stress value from the shooting method. The triggering experiment was conducted on three batches of bending units with different actuator ratios and the same dimension of $100~mm \times 7.2~mm \times 4~mm$, and the triggering environment is controlled at 80\textdegree{C}. The distance between two selected points on two ends of the triggered block is measured to represent the bending curvature and describe the bending extent.

Meanwhile, the corresponding simulation model of the bending unit is built based on the aforementioned material property definition, and the boundary conditions are shown in Figure \ref{fig:loading_unloading}(B). The initial residual stress for a particular bending unit is randomly set in Abaqus and the corresponding deformation of the block is derived after the simulation. Based on the deformation results, the difference of the measured distances between simulation and experiment can be obtained, and the initial residual stress value is subsequently adjusted to reduce this difference. Since the material's hyperelasticity is dependent on the initial residual stress value, the changing of the material property definition follows the changing of the initial residual stress. By iteratively conducting this adjustment, the initial residual stress value is identified, and the corresponding hyperelasticity definition is also finalized.

\subsection{Sequential deformation simulation}
Aside from accurate residual stress and material property identification, we propose a sequence of the deformation simulations, which can help improve the simulation accuracy especially for the cases with both the body force and the initial stress considered. Since the residual stress is identified as the initial condition for precise simulation, and we assume that the temperature during the triggering (deforming) process is a constant, the sequential deformation simulation is not time-dependent and temperature-dependent. Assume there is a model $\alpha$ to be simulated, the sequence includes (1) obtaining the initial stress value; (2) duplicating the model $\alpha$ to create a new model $\beta$, and simulating the model $\alpha$ with both the initial stress and the body force considered; (3) importing the deformation results from $\alpha$ after (2) into $\beta$ as the initial state; and (4) removing the initial stress condition in $\beta$ and simulating $\beta$ with only the body force boundary condition.

\begin{figure}[htb]
  \centering{
  \includegraphics[width=\linewidth]{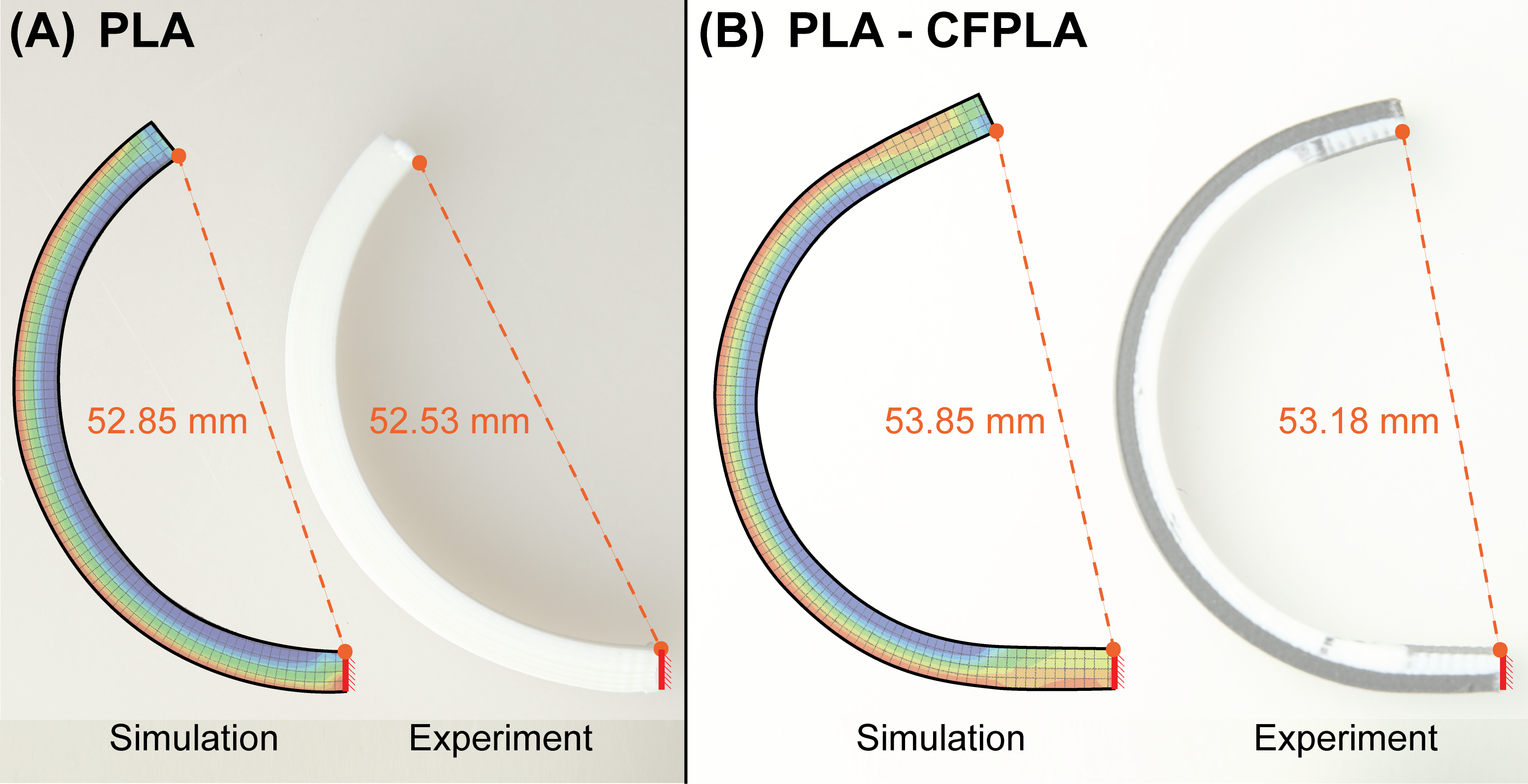}
  \caption{
    \label{fig:Beam_FEA_Deformation} Simulation and experimental results of a bending unit for PLA (A) and PLA-CFPLA composite (B).}}
\end{figure}

To verify the aforementioned FEA sequences, we trigger two bending units with different material distributions and actuator ratios - one is a PLA bending unit with the actuator ratio of $1.0$, and the other is a PLA-CFPLA bending unit with the actuator ratio of $0.75$. We use the FEA sequences to simulate these two bending units with the same geometry and material settings as experiments (the FEA model and results of the PLA bending unit are shown in Figure \ref{fig:loading_unloading}(B)). The experimental and simulation results are shown in Figure \ref{fig:Beam_FEA_Deformation}. For the PLA and PLA-CFPLA bending units, the errors between experiment and simulation are $0.61\%$ and $1.26\%$, respectively.

\textbf{Discussion 5.2.} \textit{The shooting method is used to convert the boundary value problem to an initial value problem, which can simplify the problem solving process. In this paper, we assume that the residual stress is homogeneously distributed in the sample and is released linearly, and we use the distance between each specified point pair to measure the deformation. However, the actual deforming process of the 4D printed material is very complicated. For example, the rate of residual stress releasing is not a constant of time, and the distribution of the residual stress in each sample is not homogeneous. As such, further study is required to more precisely identify the stress and control the deformation of the material. }






\section{4D printing of three creative designs} 
\label{sec:result}
Three creative designs were completed using the proposed workflow to explore potential applications. Designers can use the workflow to create 2D grid structures in a CAD environment, and then simulate the grid deformation in the Abaqus software to visualize the final shape. With our workflow, the deformation of the grid structure can be altered toward the target shape by iteratively adjusting the initial 2D grid based on the intermediate simulation results. Numerous design iterations can be conducted without prototyping, and the accuracy of the simulation - the experiment to simulation ratio in terms of deformation - is verified by implementing the point-pair method - measuring the distance between any pair of two points in both experiment and simulation. The final design can be achieved with high efficiency at low cost, and the general accuracy of the simulation, whose $95\%$ confidence interval is $(0.972, 0.985)$, can greatly assist the forward design process. 

\begin{figure}[!htb]
\center{\includegraphics[width=\linewidth]
{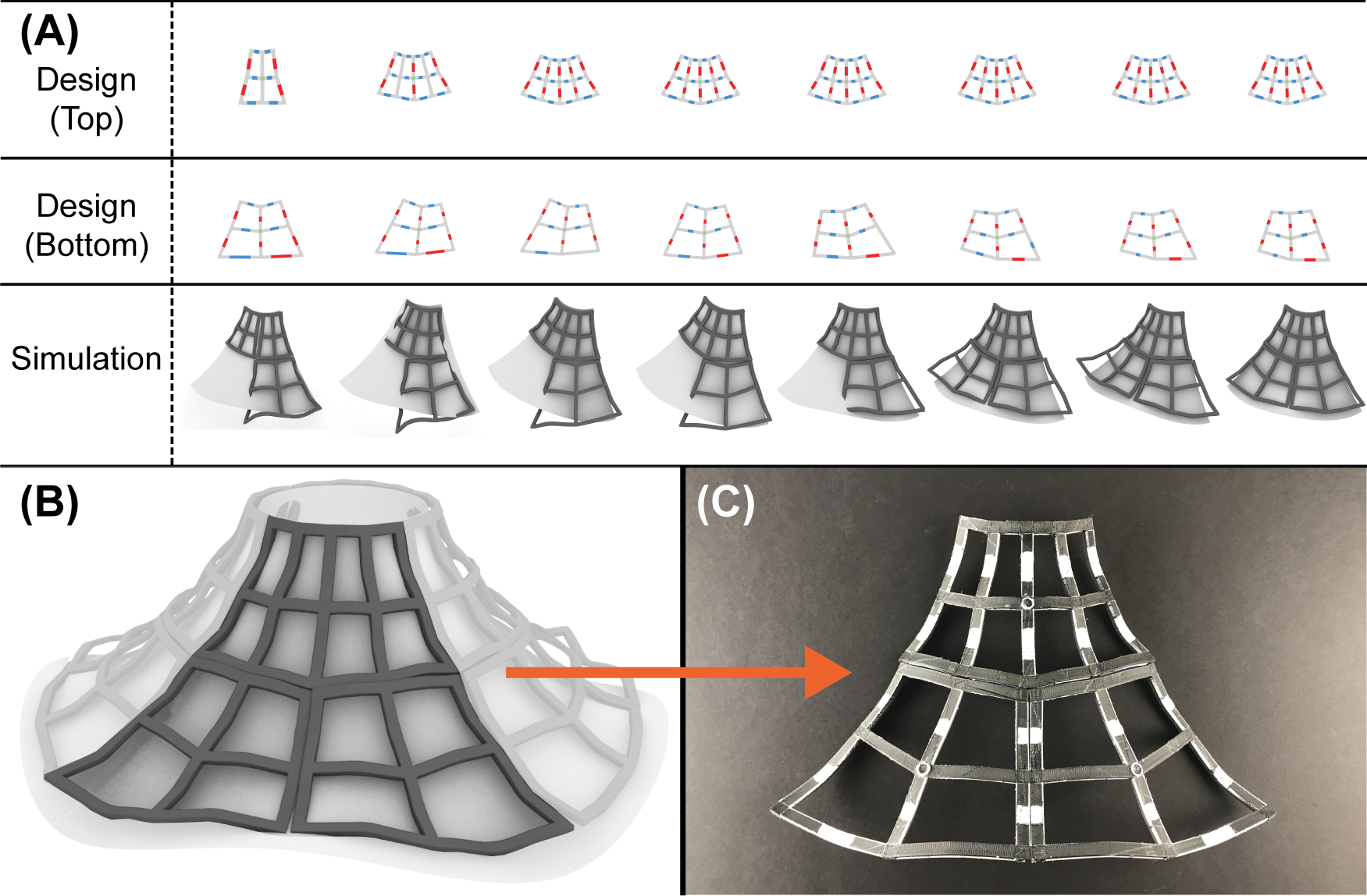}
\caption{\label{fig:Modular_Lamp_Iteration} Modular lamp cover design iterations. Results show the lamp cover pieces can fit the lamp quite well after eight iterations.}}
\end{figure}

\textbf{The modular lamp cover} is the first implementation example. The purpose of creating modular designs is to test the accuracy of our simulation tool and to allow for large scale implementation. The lamp cover is divided into three pieces: a 2x4 grid as the top section, a 2x2 grid as the bottom left section and another 2x2 grid as the bottom right section. Each sub-section design is constructed based on a 2x2 or 2x4 grid, and each grid consists of bending units and joints shown in Figures 2 and 8. By combining bending units with different actuator blocks, the 2x2 or 2x4 grid structure can achieve specific edge shapes that match each other. Once the 2D grid structures are printed, they are placed in an 80-degree Celsius water bath to trigger the deformation. This approach drastically reduces the manufacturing time because all sub-sections can be printed flat without any support structure. With the workflow, designers can iterate designs for the best matches without printing and triggering each design, significantly reducing time and cost. Figure \ref{fig:Modular_Lamp_Iteration}(A) shows the design iterations completed with our proposed workflow. Figure \ref{fig:Modular_Lamp_Iteration}(B-C) show that our simulation tool can precisely simulate the actual deformation.

\begin{figure}[!ht]
  \centering{
  \includegraphics[width=\linewidth]{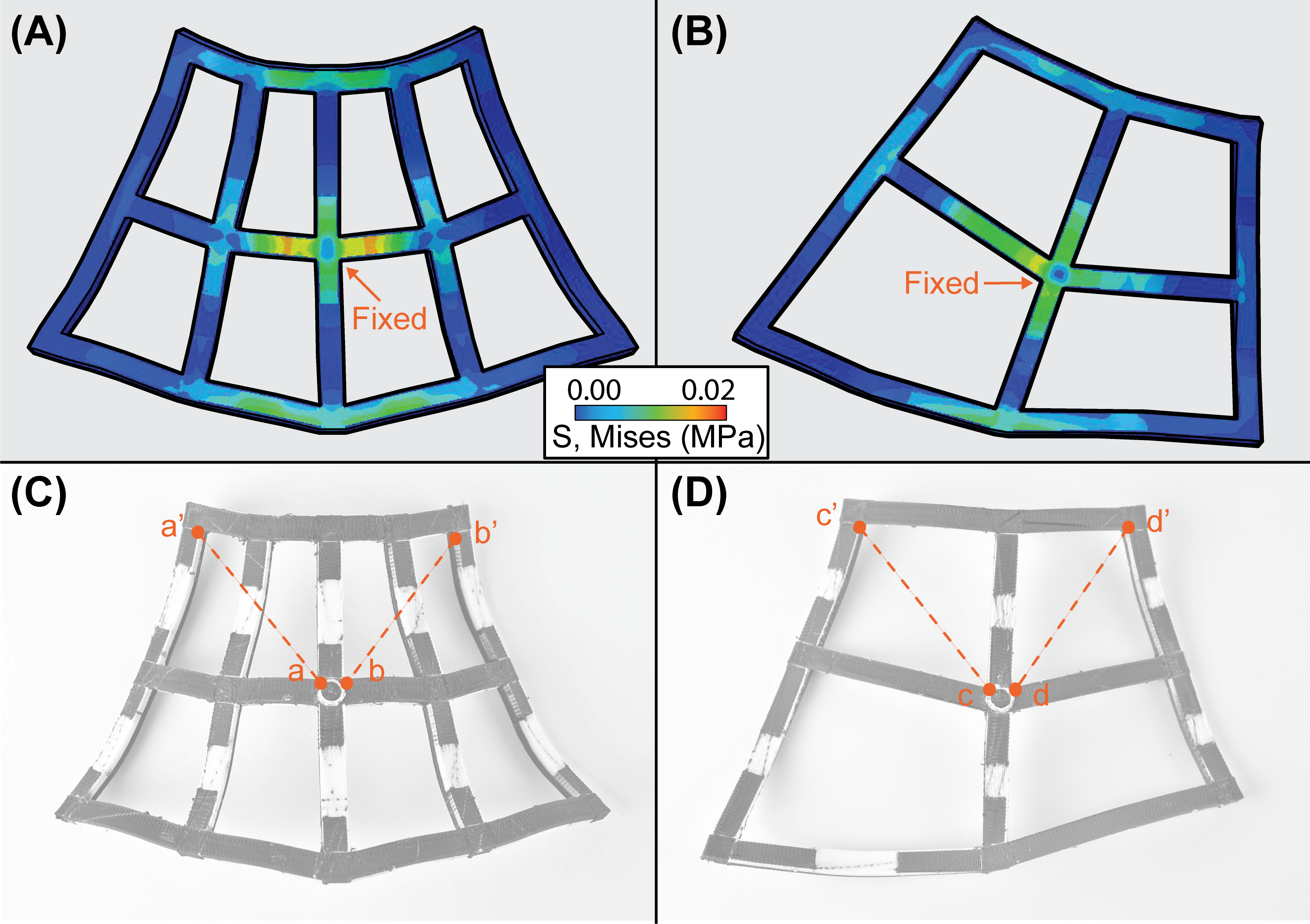}
  \caption{%
    \label{fig:lamp_deform}
    Final configurations from simulation (A-B) and experiment (C-D), where a-a', b-b', c-c' and d-d' are four different point pairs. Simulation accuracy is quantified by comparing the distance between each point pair in simulation and the corresponding distance in experiment.}}
\end{figure}

The FEA results for each lamp cover sub-section are shown in Figure \ref{fig:lamp_deform}(A-B). Since both bottom sub-sections are perfectly symmetric, only one FEA result is needed for analysis. For each subsection, three different point pairs are chosen to quantify the accuracy of the simulation results. The distance between each point pair is measured in both simulation results and experimental results. Then the simulation error is determined by calculating the difference between experimental and simulation distances. Part of the distance results and the calculated errors are listed in Table \ref{tab:lamp_accuracy}. More measurement results are provided in the supplementary material (Table S4). Based on the aforementioned measurement results, the $95\%$ confidence interval for the accuracy is $(0.968, 0.998)$.

\begin{small}
\begin{table}[!ht]
    \centering{
    \caption{Simulation-experiment error for the lamp cover}
    \vspace{-2mm}
    \begin{tabular}{c|c|c|c}
        \hline
        Point pairs & Experiment (mm) & Simulation (mm) & Error (\%) \\
        \hline
        a-a' & 76.54 & 72.42 & 5.38 \\ 
        b-b' & 78.16 & 78.50 & 0.44 \\ 
        c-c' & 64.95 & 66.01 & 1.63 \\ 
        d-d' & 65.06 & 61.93 & 4.81 \\ 
        \hline
    \end{tabular}
    \label{tab:lamp_accuracy}}
\end{table}
\end{small}

\begin{figure}[!htb]
\center{\includegraphics[width=\linewidth]
{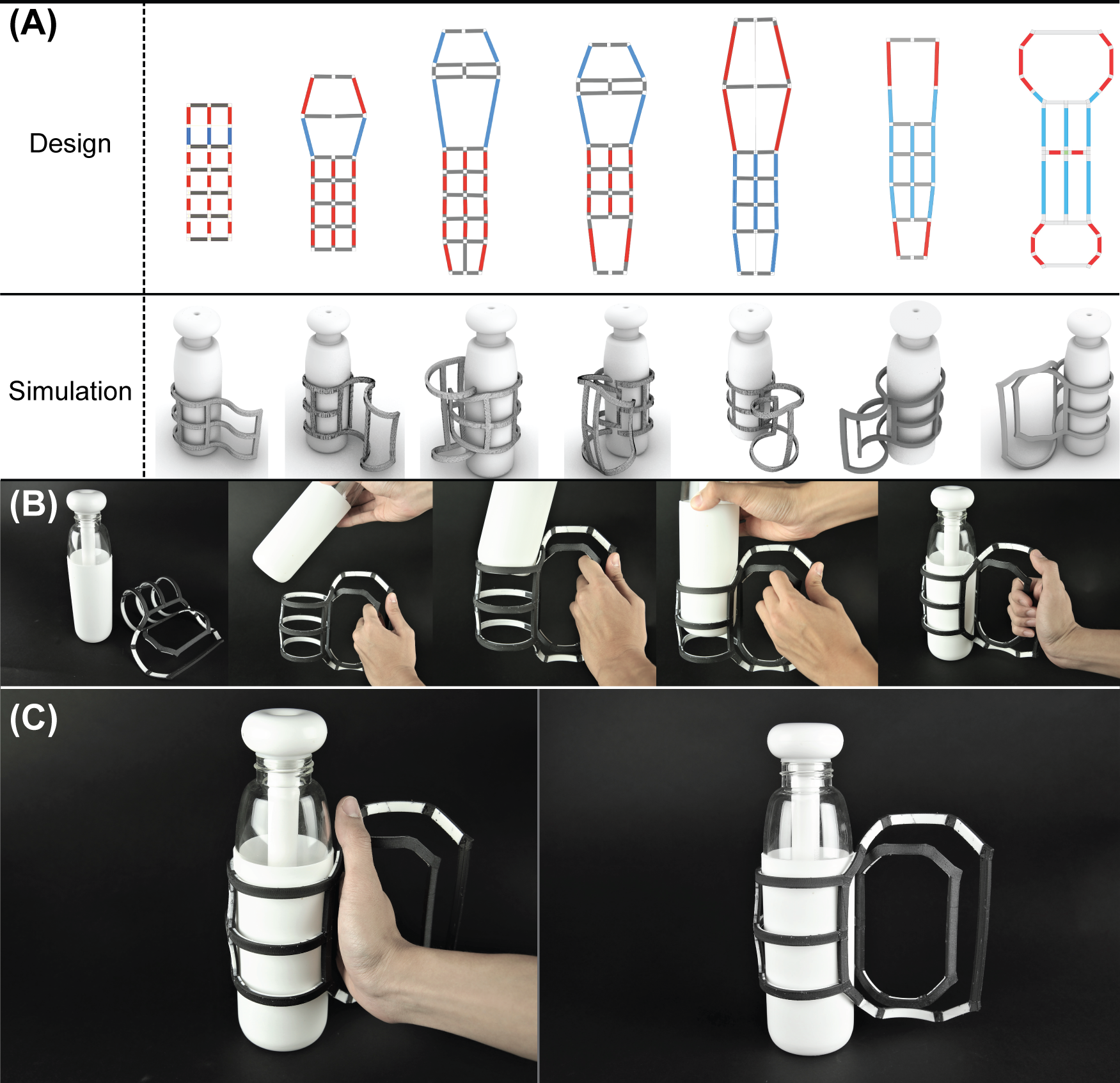}
\caption{\label{fig:Bottle_Holder_Iteration} Bottle holder design iteration. Results show the bottle holder can fit the bottle quite well after six iterations.}}
\end{figure}

\textbf{The bottle holder} is the second implementation example of our workflow. The purpose of designing a bottle holder is to enable people with disabilities to hold bottles without handles. The bottle holder is an evolved version of a 2x2 grid structure with an approximate dimension of 500 mm by 200 mm. By changing the construction of the 2D grid, complex and intersecting 3D structures like the bottle holder can be realized. With the conventional 3D printing strategy shown in Figure \ref{fig:Bottle_Holder_Iteration}, the bottle holder structure will require extensive support structures and extended printing time. With the proposed workflow, designers can iterate through different designs and achieve the best fit for a specific bottle with significantly faster speed. Once the design is finalized, FDM printed 2D grid structure is printed and placed in an 80\textdegree{C} water bath. The residual stress embedded in the material and the gravity will deform the grid into the programmed bottle holder shape.

\begin{figure}[!htb]
\center{\includegraphics[width=0.85\linewidth]
{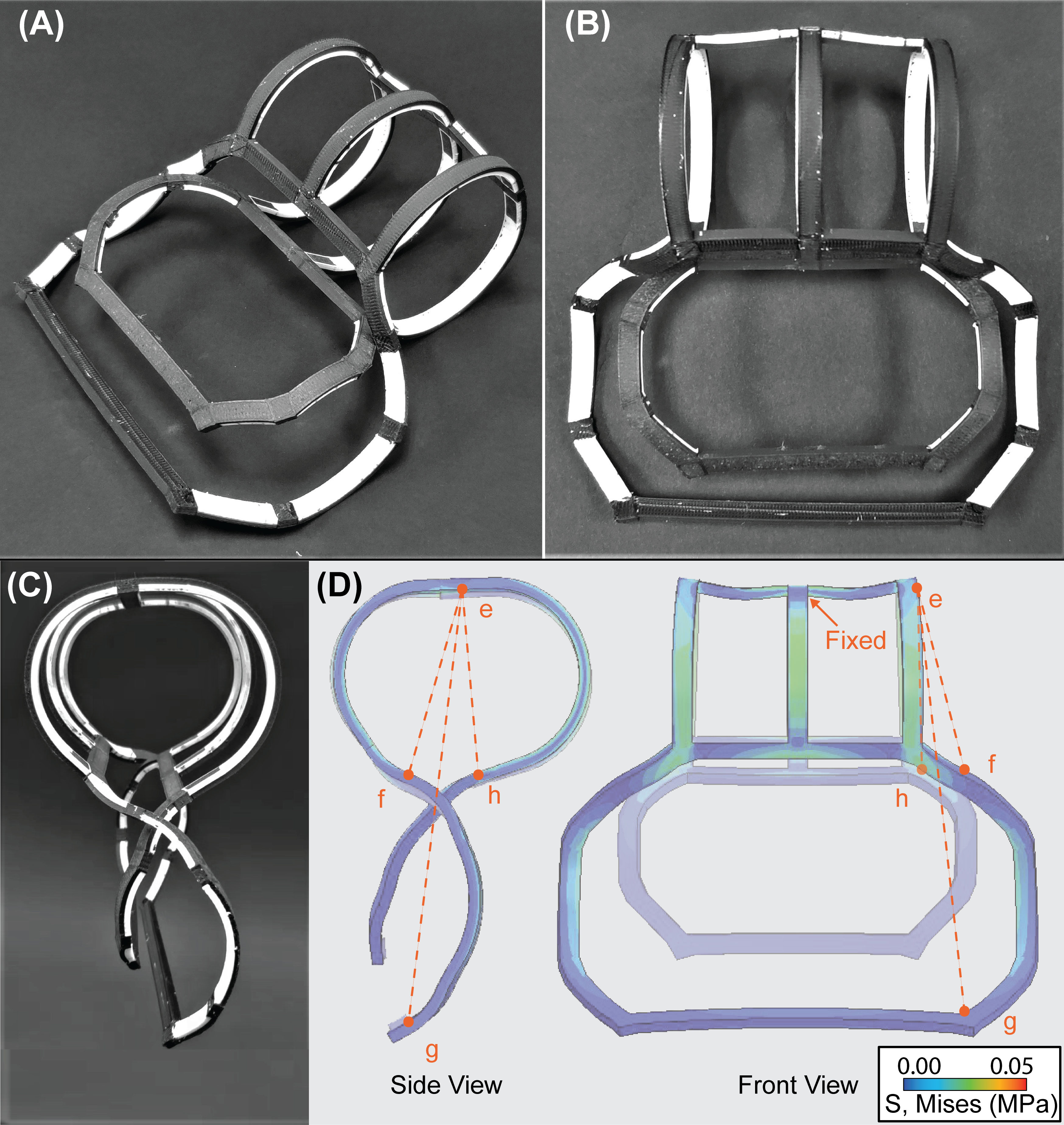}
\caption{\label{fig:Composite_PLA_Bottle_Holder} Composite bottle holder. (A), (B) and (C) are different views of the final triggered result. (D) is the simulation result of the final configuration, where the labeled point pairs are used for the measurement and the accuracy calculation.}}
\end{figure}

The deformation result is shown in Figure \ref{fig:Composite_PLA_Bottle_Holder}(A-C). The error between simulation and experiment is also calculated based on distance measurements. Three point pairs on the bottle holder are labeled in Figure \ref{fig:Composite_PLA_Bottle_Holder}(D). Table \ref{tab:bottle_accuracy} shows part of the distance results for each point pair and the corresponding error. More measurement results are provided in the supplementary material (Table S4). Based on the aforementioned measurement results, the $95\%$ confidence interval for the accuracy is $(0.962, 0.986)$.

\begin{small}
\begin{table}[!ht]
    \centering{
    \caption{Simulation-experiment error of the bottle holder}
    \vspace{-2mm}
    \begin{tabular}{c|c|c|c}
        \hline
        Point pairs & Experiment (mm) & Simulation (mm) & Error (\%) \\
        \hline
        e-f & 71.35 & 70.68 & 0.94 \\ 
        e-g & 161.45 & 160.31 & 0.43 \\ 
        e-h & 67.58 & 64.98 & 3.85 \\ 
        \hline
    \end{tabular}
    \label{tab:bottle_accuracy}}
\end{table}
\end{small}

\textbf{The shoe supporter} is the third designed product. This application involves adapting sneakers for high-performance contexts such as hiking or climbing. This design is an exploration of how the composite material can interface with both the existing products and the human body. Employing the same strategy as the bottle holder, this design consists of two pieces - the top part and the bottom part. The design relies on connecting these two pieces attached to the sneaker and the ankle using elastic straps that anchor the bottom piece and strategically limit the movement of the top piece. In its active state, this design is meant to prevent injuries such as a rolled or sprained ankle, as well as to provide additional protection on the foot and the ankle in harsh conditions. 

\begin{figure}[!htb]
\center{\includegraphics[width=\linewidth]
{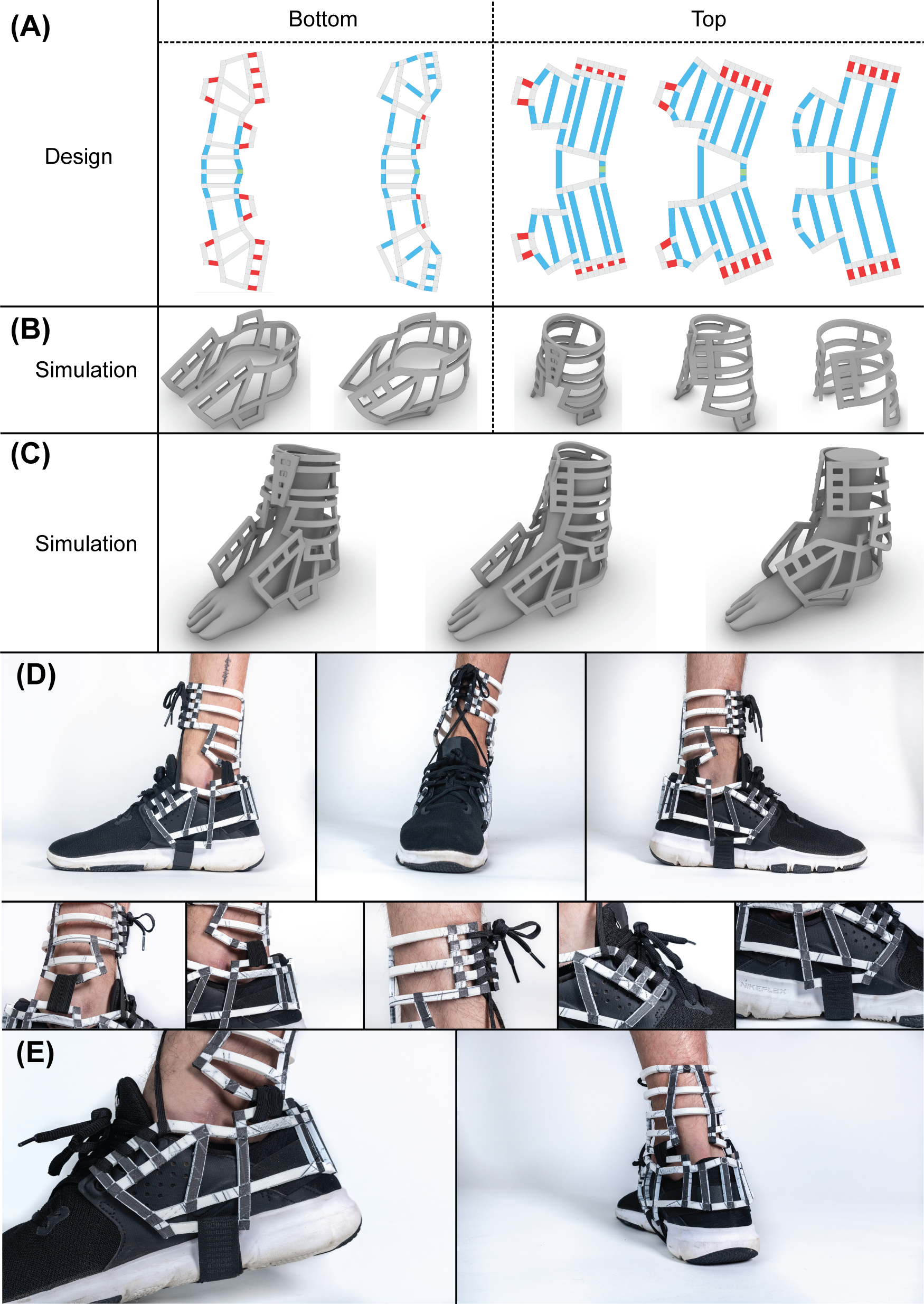}
\caption{\label{fig:Shoe_Support_Design_Iteration}The design iteration of the shoe supporter. The shape of the shoe supporter design is finalized to fit the shape of the foot by going through this iterative process.}}
\end{figure}

\begin{figure}[!htb]
\center{\includegraphics[width=\linewidth]
{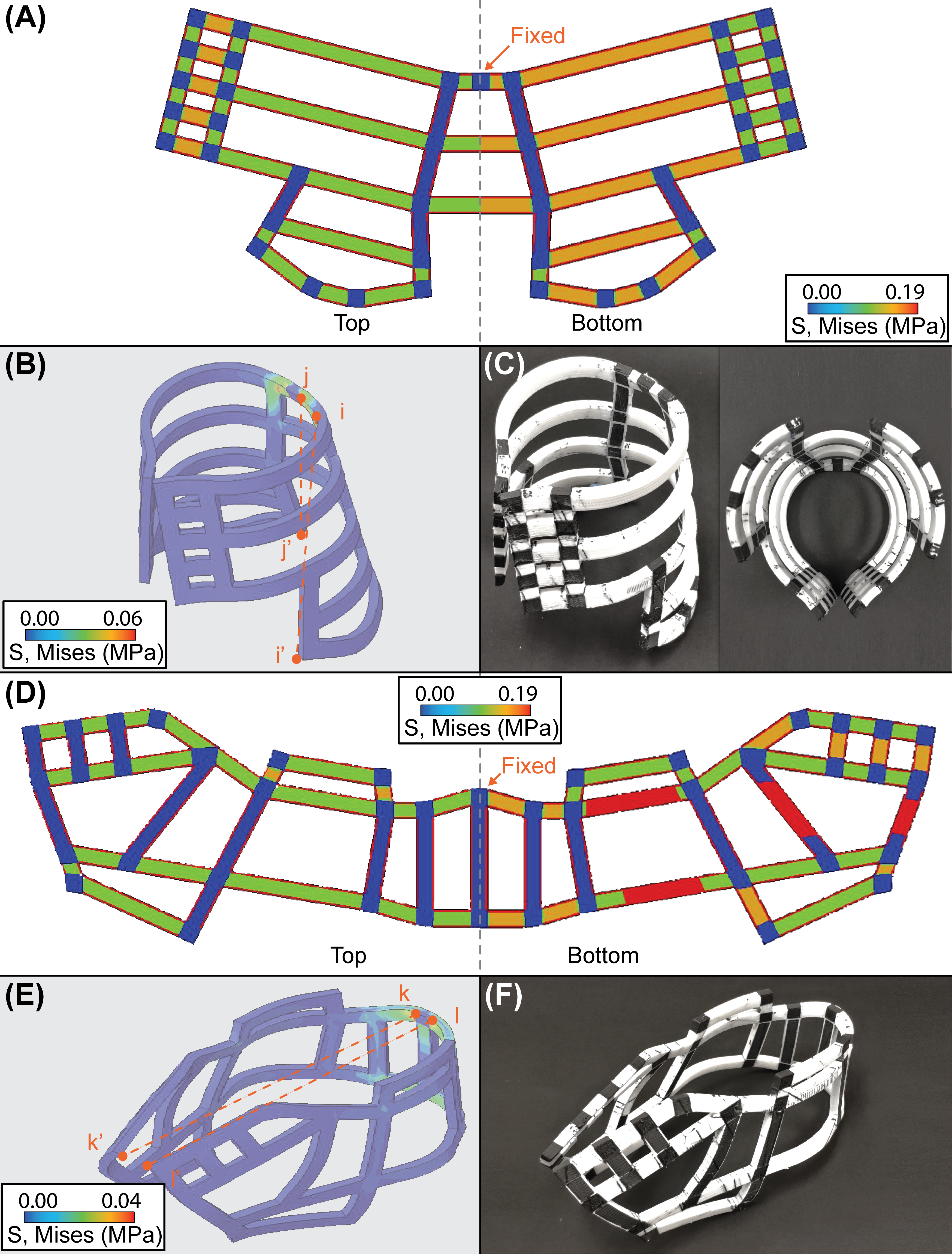}
\caption{\label{fig:Shoe_Support_Simulation} The simulation and experimental results of the shoe supporter with the top piece (A-C) and the bottom piece (D-F). (A, D) show the initial condition from both the top view and the bottom view. (B, E) show the simulation result, where the labeled point pairs are used for the measurement and the accuracy calculation. (C, F) show the experimental results.}}
\end{figure}

FEA results of both the top piece and the bottom piece are shown in Figure \ref{fig:Shoe_Support_Design_Iteration}. Experimental results of the finalized sample are shown in Figure \ref{fig:Shoe_Support_Simulation}. The error is measured using the same strategy as the lamp cover and the bottle holder. Four point pairs are labeled in Figure \ref{fig:Shoe_Support_Simulation}(B, E). Part of the measurement results and the calculated error are listed in Table \ref{tab:shoe_accuracy}. More measurement results are provided in the supplementary material (Table S4). Based on the aforementioned measurement results, the $95\%$ confidence interval of the accuracy is $(0.969, 0.988)$.

\begin{small}
\begin{table}[!ht]
    \centering{
    \caption{Simulation-experiment error of the shoe support}
    \vspace{-2mm}
    \begin{tabular}{c|c|c|c}
        \hline
        Point pairs & Experiment (mm) & Simulation (mm) & Error (\%) \\
        \hline
        i-i' & 89.74 & 92.25 & 2.80 \\ 
        j-j' & 89.83 & 92.25 & 2.69 \\ 
        k-k' & 160.57 & 157.02 & 2.21 \\ 
        l-l' & 161.86 & 157.02 & 2.99 \\ 
        \hline
    \end{tabular}
    \label{tab:shoe_accuracy}}
\end{table}
\end{small}

\section{Conclusion and future work}
\label{sec:conclusion}
In this paper, we established a computational workflow that uses FEA to produce physically accurate results of the residual stress-induced morphing behaviors of mesh-like thermoplastic composite structures. To accurately conduct the simulation, experiments have been conducted to quantify the hyperelasticity and viscoelasticity of the thermoplastic component of FRC. These tests include uniaxial tensile and compression tests for the elastic component, DMA for the viscous component and uniaxial unloading-reloading with different strain levels for plastic deformation and the Mullins effect. This paper has also introduced the simulation sequence based on the triggering experiment and estimated the residual stress. The paper has validated the FEA modeling with an accurate matching between simulation and experiment for three creative designs.

One of the main limitations of the presented workflow is the high computational cost of FEA simulations, which makes it impossible for real-time, interactive design. Here we consider Graph Neural Network (GNN) as a potential model to speed up the simulation. It is worth noticing that the underlying topology of the grid structure can be abstracted to an undirected weight graph that is well compatible with GNN \cite{zhou2018graph}. We are excited to look for appropriate abstraction and simplification by the aid of machine learning, and to explore a simulation method with both mechanical accuracy and computational efficiency.

\section*{Acknowledgment}
This project was mainly supported by the Carnegie Mellon University Manufacturing Futures Initiative (MFI) made possible by the Richard King Mellon Foundation, and a Honda grant. We are also especially thankful to Mr. William Pingitore, who supervised us conducting the four-point bending experiment, and Dr. Guanyun Wang, who helped us on multiple design applications, and Prof. Christopher Bettinger in the Departments of Materials Science and Engineering and Biomedical Engineering, who provided the DMA facility. The authors acknowledge the use of the Materials Characterization Facility at Carnegie Mellon University supported by the grant MCF-677785.

\section*{References}
\bibliographystyle{elsarticle-num-names}
\bibliography{pla_paper}

\end{document}


\setcounter{section}{0}
\renewcommand{\thesection}{S\arabic{section}}
\setcounter{table}{0}
\renewcommand{\thetable}{S\arabic{table}}
\setcounter{figure}{0}
\renewcommand{\thefigure}{S\arabic{figure}}
\part*{Supplementary material}
\section{Viscoelasticity \& hyperelasticity data for ABAQUS}
This section includes the hyperelastic and viscoelastic data used in ABAQUS models. The main loading stress-strain data is provided in Table~\ref{tab:Loading_curve_main}. As shown in Figure 9, this curve is used to obtain several unloading stress-strain curves based on different residual stress values. Several unloading stress-strain data of hyperelasticity with corresponding residual stress values are provided in Table~\ref{tab:Unloading_stres_strain}, which can be used as hyperelasticity definition data for ABAQUS.
For more information about the hyperelasticity definition based on the unloading stress-strain curve, please refer to Section 5 in the main paper.

\begin{table}[!ht]
    \centering{
    \caption{Loading stress-strain data}
    \vspace{-2mm}
    \begin{tabular}{r|r}
        \hline
        Engineering Strain & Engineering Stress (MPa) \\
        \hline
        0 & 0  \\
        0.000232 & 0.00060322 \\
        0.000464 & 0.00120614 \\
        0.000812 & 0.00210997 \\
        0.001334 & 0.00346446 \\
        0.002117 & 0.00549213 \\
        0.0032915 & 0.00851022 \\
        0.00505325 & 0.0122125 \\
        0.00769588 & 0.0173997 \\
        0.0116598 & 0.0247775 \\
        0.0176057 & 0.0349489 \\
        0.0265246 & 0.0487535 \\
        0.0381246 & 0.0647783 \\
        0.0497246 & 0.0793461 \\
        0.0613246 & 0.0925438 \\
        0.0729246 & 0.104502 \\
        0.0845246 & 0.115441 \\
        0.0961246 & 0.125616 \\
        0.107725 & 0.134868 \\
        0.119325 & 0.143482 \\
        0.130925 & 0.151579 \\
        0.142525 & 0.158984 \\
        0.154125 & 0.166053 \\
        0.165725 & 0.172959 \\
        0.177325 & 0.179292 \\
        0.188925 & 0.185297 \\
        0.200525 & 0.190677 \\
        0.212125 & 0.195992 \\
        0.223725 & 0.200807 \\
        0.232 & 0.204067 \\
        \hline
    \end{tabular}
    \label{tab:Loading_curve_main}}
\end{table}

\newcommand{\tabincell}[2]{\begin{tabular}{@{}#1@{}}#2\end{tabular}}
\begin{table*}[!ht]
    \centering{
    \caption{Unloading stress-strain data with different initial residual stresses}
    \vspace{-2mm}
    \begin{tabular}{c|c|c|c|c|c|c|c}
        \hline
        \multicolumn{2}{c|}{\tabincell{c}{Initial residual stress \\ 0.203 MPa}} & \multicolumn{2}{c|}{\tabincell{c}{Initial residual stress \\ 0.170 MPa}} & \multicolumn{2}{c|}{\tabincell{c}{Initial residual stress \\ 0.132 MPa}} & \multicolumn{2}{c}{\tabincell{c}{Initial residual stress \\ 0.079 MPa}} \\
        \hline
        \tabincell{c}{Engineering \\ Strain} & \tabincell{c}{Engineering \\ Stress (MPa)} & \tabincell{c}{Engineering \\ Strain} & \tabincell{c}{Engineering \\ Stress (MPa)} & \tabincell{c}{Engineering \\ Strain} & \tabincell{c}{Engineering \\ Stress (MPa)} & \tabincell{c}{Engineering \\ Strain} & \tabincell{c}{Engineering \\ Stress (MPa)} \\
        \hline
        0.231577 & 0.203041 & 0.162681 & 0.170461 & 0.104322 & 0.132095 & 0.049452 & 0.079267 \\
0.231181 & 0.202518 & 0.162414 & 0.170022 & 0.104158 & 0.131755 & 0.049392 & 0.079063 \\
0.230648 & 0.201734 & 0.162119 & 0.169363 & 0.103963 & 0.131244 & 0.049271 & 0.078757 \\
0.22985 & 0.200557 & 0.161635 & 0.168375 & 0.103624 & 0.130479 & 0.049082 & 0.078297 \\
0.228679 & 0.198792 & 0.160825 & 0.166893 & 0.103107 & 0.12933 & 0.048861 & 0.077608 \\
0.226846 & 0.196144 & 0.159674 & 0.16467 & 0.102343 & 0.127608 & 0.04846 & 0.076574 \\
0.224179 & 0.192172 & 0.157945 & 0.161336 & 0.101161 & 0.125024 & 0.047888 & 0.075024 \\
0.22009 & 0.186215 & 0.155266 & 0.156335 & 0.099355 & 0.121148 & 0.047004 & 0.072698 \\
0.213945 & 0.177279 & 0.151203 & 0.148833 & 0.096469 & 0.115335 & 0.04567 & 0.069209 \\
0.207032 & 0.167076 & 0.146264 & 0.140267 & 0.093033 & 0.108697 & 0.044045 & 0.065226 \\
0.200133 & 0.156873 & 0.141333 & 0.1317 & 0.089435 & 0.102059 & 0.042368 & 0.061242 \\
0.193013 & 0.146669 & 0.136189 & 0.123134 & 0.085708 & 0.09542 & 0.040592 & 0.05726 \\
0.185701 & 0.136466 & 0.130879 & 0.114568 & 0.081911 & 0.088783 & 0.038742 & 0.053276 \\
0.178143 & 0.126263 & 0.125469 & 0.106002 & 0.07793 & 0.082144 & 0.036825 & 0.049293 \\
0.170048 & 0.116059 & 0.119661 & 0.097436 & 0.073899 & 0.075506 & 0.034817 & 0.045309 \\
0.161563 & 0.105856 & 0.113494 & 0.08887 & 0.069693 & 0.068868 & 0.032716 & 0.041326 \\
0.152897 & 0.095653 & 0.106497 & 0.080304 & 0.065287 & 0.06223 & 0.0305 & 0.037343 \\
0.14397 & 0.085449 & 0.099488 & 0.071738 & 0.06077 & 0.055592 & 0.028206 & 0.033359 \\
0.134307 & 0.075245 & 0.091882 & 0.063172 & 0.056154 & 0.048954 & 0.025726 & 0.029376 \\
0.123376 & 0.065043 & 0.084797 & 0.054606 & 0.050867 & 0.042316 & 0.023132 & 0.025393 \\
0.112395 & 0.054839 & 0.076873 & 0.04604 & 0.045547 & 0.035677 & 0.020671 & 0.021409 \\
0.101224 & 0.044636 & 0.068423 & 0.037473 & 0.039771 & 0.02904 & 0.017677 & 0.017426 \\
0.089966 & 0.034432 & 0.060173 & 0.028907 & 0.033847 & 0.022401 & 0.01498 & 0.013442 \\
0.078535 & 0.024229 & 0.051336 & 0.020341 & 0.0279 & 0.015763 & 0.011912 & 0.009459 \\
0.067843 & 0.014025 & 0.043237 & 0.011775 & 0.022177 & 0.009125 & 0.008823 & 0.005476 \\
0.058737 & 0.003823 & 0.035832 & 0.003209 & 0.01703 & 0.002487 & 0.006036 & 0.001492 \\
0.055328 & 0.0 & 0.03359 & 0.0 & 0.015219 & 0.0 & 0.004998 & 0.0 \\
        \hline
    \end{tabular}
    \label{tab:Unloading_stres_strain}}
\end{table*}

The viscoelasticity definition data for ABAQUS is provided in Table~\ref{tab:visco}. These data are directly from the DMA experiments. Each row defines the real and imaginary parts of Young's modulus (storage and loss modulus) at corresponding frequency. The frequency describes cycles of applied loading per time. The uniaxial strain shows the preloaded shape where a cyclic loading is applied to.

\begin{table*}[!ht]
    \centering{
    \caption{The modulus-frequency data at pre-strains 0.0095 from 0.01 Hz to 100 Hz for PLA and CFPLA.}
    \vspace{-2mm}
    \begin{tabular}{c|c|c|c|c|c|c|c|c|c}
    \hline
        \multicolumn{5}{c|}{\tabincell{c}{PLA}} & \multicolumn{5}{c}{\tabincell{c}{CFPLA}} \\
        \hline
        \tabincell{c}{Loss \\ Modulus \\ (MPa)} & \tabincell{c}{Storage \\ Modulus \\ (MPa)} & \tabincell{c}{$tan \delta$} & \tabincell{c}{Frequency \\ (Hz)} & \tabincell{c}{Uniaxial \\ Strain} & \tabincell{c}{Loss \\ Modulus \\ (MPa)} & \tabincell{c}{Storage \\ Modulus \\ (MPa)} & \tabincell{c}{$tan \delta$} & \tabincell{c}{Frequency \\ (Hz)} & \tabincell{c}{Uniaxial \\ Strain} \\
        \hline
        0.620318 & 2.49694 & 0.248431 & 0.01 & 0.0095 & 10.2519 & 46.9742 & 0.218246 & 0.01 & 0.0095 \\
0.673856 & 3.23405 & 0.208363 & 0.031623 & 0.0095 & 18.4361 & 117.366 & 0.157082 & 0.039811 & 0.0095 \\
0.787532 & 3.79361 & 0.207594 & 0.1 & 0.0095 & 23.9765 & 159.709 & 0.150126 & 0.1 & 0.0095 \\
0.914541 & 4.22998 & 0.216205 & 0.158489 & 0.0095 & 29.3856 & 186.844 & 0.157273 & 0.199526 & 0.0095 \\
0.992119 & 4.44871 & 0.223013 & 0.251189 & 0.0095 & 31.3125 & 194.104 & 0.161319 & 0.251189 & 0.0095 \\
1.19299 & 4.54469 & 0.262502 & 0.398107 & 0.0095 & 37.1118 & 206.143 & 0.180029 & 0.398107 & 0.0095 \\
1.2871 & 5.12296 & 0.251241 & 0.630957 & 0.0095 & 39.9765 & 219.115 & 0.182445 & 0.630957 & 0.0095 \\
1.51195 & 5.44861 & 0.277493 & 1.0 & 0.0095 & 47.3401 & 231.911 & 0.204131 & 1.0 & 0.0095 \\
1.6472 & 5.64237 & 0.291934 & 1.25893 & 0.0095 & 51.9323 & 239.84 & 0.216529 & 1.25893 & 0.0095 \\
1.80734 & 5.85958 & 0.308442 & 1.58489 & 0.0095 & 61.6246 & 256.361 & 0.240382 & 1.99526 & 0.0095 \\
1.99242 & 6.08222 & 0.327581 & 1.99526 & 0.0095 & 67.6072 & 265.431 & 0.254707 & 2.51189 & 0.0095 \\
2.20423 & 6.30915 & 0.34937 & 2.51189 & 0.0095 & 74.3216 & 275.296 & 0.26997 & 3.16228 & 0.0095 \\
2.40932 & 6.58269 & 0.366008 & 3.16228 & 0.0095 & 90.1077 & 298.653 & 0.301714 & 5.01187 & 0.0095 \\
2.75754 & 6.85159 & 0.402467 & 3.98107 & 0.0095 & 99.9106 & 312.204 & 0.320018 & 6.30957 & 0.0095 \\
3.05346 & 7.16496 & 0.426166 & 5.01187 & 0.0095 & 110.755 & 328.133 & 0.337532 & 7.94328 & 0.0095 \\
3.47829 & 7.45562 & 0.466533 & 6.30957 & 0.0095 & 122.348 & 348.268 & 0.351304 & 10.0 & 0.0095 \\
3.94031 & 7.96939 & 0.494431 & 7.94328 & 0.0095 & 145.115 & 388.692 & 0.373342 & 14.4544 & 0.0095 \\
4.45309 & 8.31036 & 0.535848 & 10.0 & 0.0095 & 164.059 & 423.77 & 0.387141 & 19.0546 & 0.0095 \\
4.85071 & 8.79196 & 0.551721 & 12.5893 & 0.0095 & 178.334 & 450.418 & 0.395929 & 22.9087 & 0.0095 \\
5.7382 & 9.30525 & 0.616663 & 15.8489 & 0.0095 & 192.897 & 479.298 & 0.402458 & 27.5423 & 0.0095 \\
6.69739 & 9.7508 & 0.686855 & 19.9526 & 0.0095 & 214.785 & 528.393 & 0.406487 & 36.3078 & 0.0095 \\
7.57122 & 10.4126 & 0.727121 & 25.1189 & 0.0095 & 234.739 & 587.395 & 0.399628 & 47.863 & 0.0095 \\
8.71075 & 11.1136 & 0.783792 & 31.6228 & 0.0095 & 240.298 & 609.785 & 0.394071 & 52.4807 & 0.0095 \\
10.1849 & 11.9006 & 0.855831 & 39.8107 & 0.0095 & 246.22 & 654.085 & 0.376435 & 63.0957 & 0.0095 \\
11.8424 & 12.7682 & 0.927492 & 50.1187 & 0.0095 & 247.398 & 678.274 & 0.364747 & 69.1831 & 0.0095 \\
13.6883 & 13.4669 & 1.01644 & 63.0957 & 0.0095 & 241.281 & 735.224 & 0.328173 & 83.1764 & 0.0095 \\
16.5182 & 15.0428 & 1.09808 & 79.4328 & 0.0095 & 223.761 & 759.381 & 0.294663 & 91.2011 & 0.0095 \\
20.2038 & 16.7091 & 1.20915 & 100.0 & 0.0095 & 222.699 & 768.417 & 0.289815 & 100.0 & 0.0095 \\
        \hline
    \end{tabular}
    \label{tab:visco}}
\end{table*}

\section{CFPLA - viscoelasticity evidence}
To evidence the decision of not considering the viscoelasticity of the CFPLA, correlated data and diagrams are provided in this section. CFPLA's storage modulus and loss modulus under different loading frequencies are obtained through the DMA experiment, and the damping ratios (loss modulus to storage modulus, which can also be abbreviated as \textit{tan(delta)}) are also calculated based on the testing results.

\begin{figure}[!ht]
    \centering
    \includegraphics[width=\linewidth]{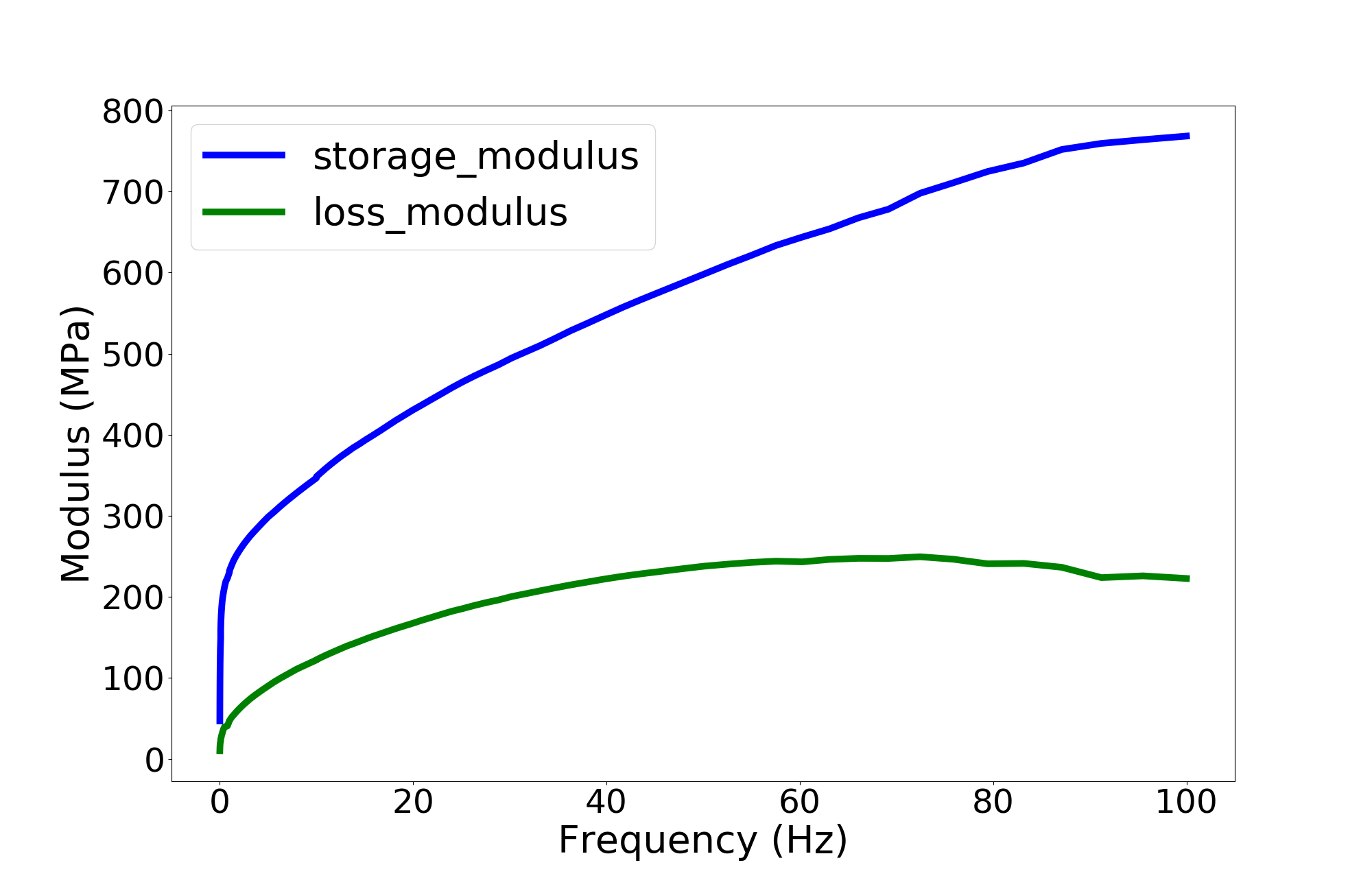}
    \caption{Storage modulus and Loss modulus of CFPLA. The curve of the storage modulus lies above the curve of the loss modulus through the whole frequency range. }
    \label{fig:CFPLA_visco_modulus}
\end{figure}

\begin{figure}[!ht]
    \centering
    \includegraphics[width=\linewidth]{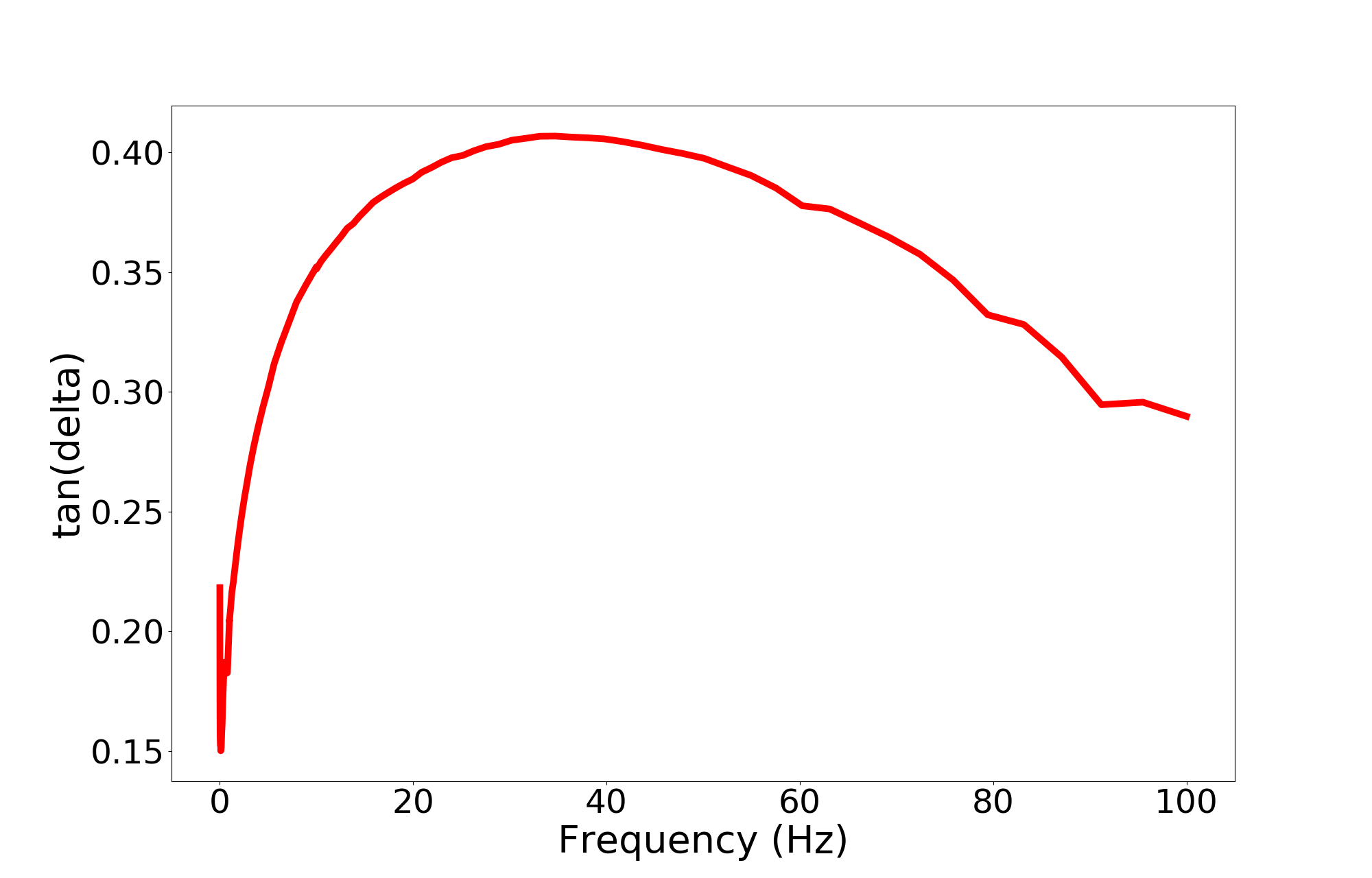}
    \caption{Damping ratio of CFPLA. Together with Figure \ref{fig:CFPLA_visco_modulus}, it can offer insight into the material physics of CFPLA that elastic properties and elastic behaviors take the dominance over viscous properties and viscous behaviors. }
    \label{fig:CFPLA_visco_tan}
\end{figure}

Table \ref{tab:visco} shows the DMA experimental results and the calculated \textit{tan(delta)} values, and Figures \ref{fig:CFPLA_visco_modulus} and \ref{fig:CFPLA_visco_tan} show the modulus curves and the damping ratio curve, respectively. It can be clearly observed that there is no intersection between the curve of storage modulus and the curve of loss modulus, and the value of \textit{tan(delta)} is less than $1.0$ through the whole range of testing frequencies. This indicates the dominance of CFPLA's elastic property and elastic behavior. As such, we reasonably abandon the CFPLA's viscoelasticity definition in our simulation model and regard the CFPLA as one type of hyperelastic material.

\section{Four-point bending experiment}
\begin{figure}[!ht]
    \centering{
    \includegraphics[width=\linewidth]{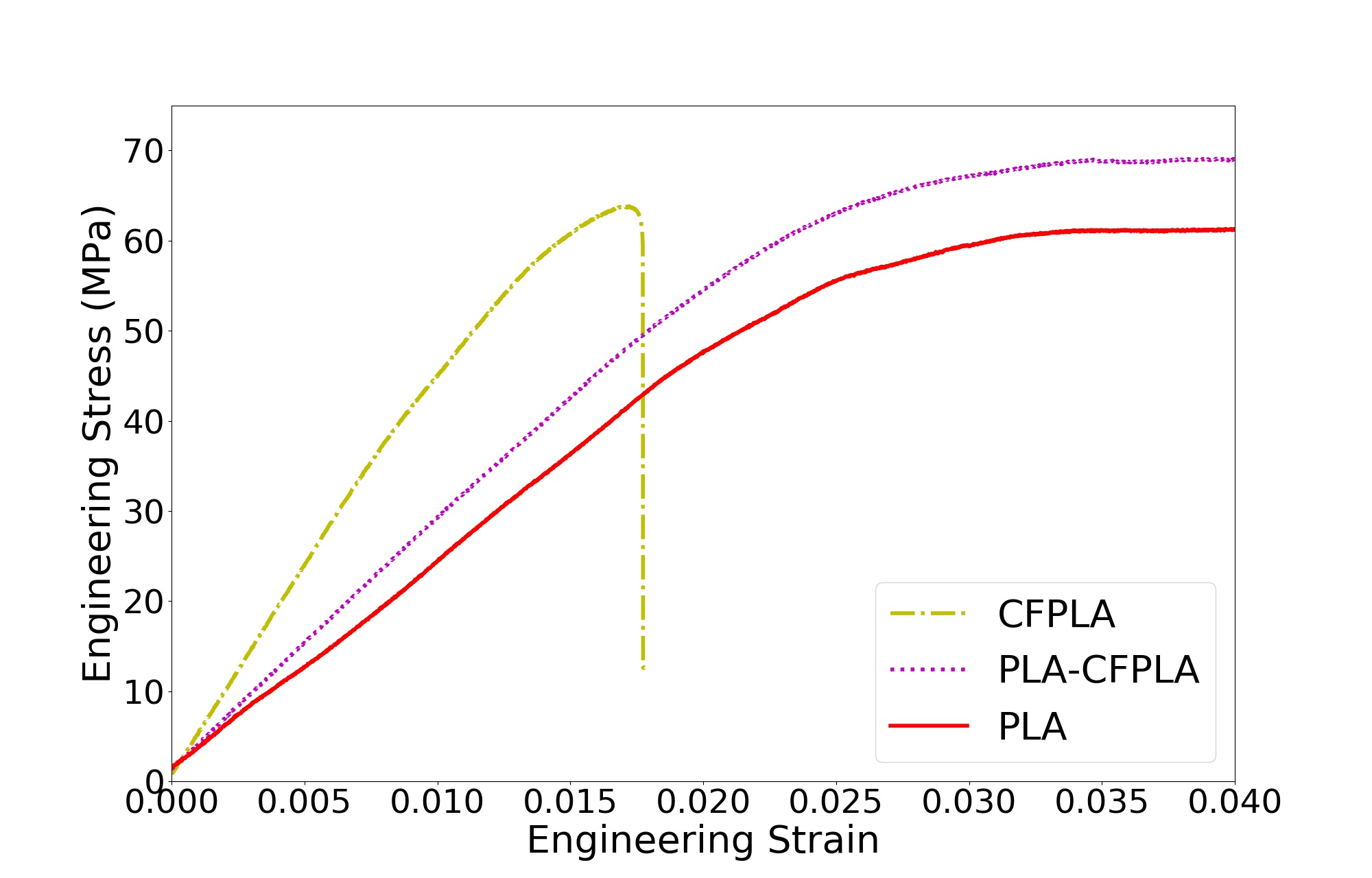}
    \caption{The engineering stress-strain curve derived from the four-point flexural test. The result indicates that CFPLA has the highest stiffness and PLA has the lowest stiffness, which is consistent with the conclusion drawn from our three-point flexural test. The four-point flexural test can also show the information of the flexural strength, the effective deflection range, and the fracture performance. }
    \label{fig:4pb}}
\end{figure}

In Section 3.3 of the paper, we present the process and results of the three-point flexural test. However, due to the limited maximum load of the DMA equipment, the material's strength and fracture performance under large deflections cannot be accurately characterized by our three-point bending setup. Moreover, since the three-point flexural test may result in a sharp curvature at the loading point, the maximum stress point will locate at the mid-point of the specimen, which is not reliable in evaluating material strength and fracture, especially for brittle composites like CFPLA. To resolve these problems and provide more solid evidence for the material's flexural performances, we additionally conduct the four-point flexural test on the same three types of structures (i.e., PLA, CFPLA, PLA-CFPLA bi-layer composite). The results of the four-point bending test are depicted in Figure \ref{fig:4pb}.

The brand of the four-point bending equipment is \textit{MTS Corp}. The dimension of each specimen for four-point bending test is $100~mm \times 8~mm \times 8~mm$, and the loading span and the supporting span are $23.56~mm$ and $70.68~mm$, respectively. The test is conducted under room temperature of 20~\textdegree{C}.

From the slope comparison in Figure \ref{fig:4pb}, we can clearly observe that the order of the stiffness of three types of structures should be (from large to small): \textbf{CFPLA $>$ PLA-CFPLA $>$ PLA}, which is consistent with the result of the three-point bending experiment. In terms of the flexural strength and the effective deflection range, CFPLA cannot withstand large deflection and has the fracture strength of 63.801~MPa, while structures with PLA have larger flexural strength than CFPLA and can withstand deflection larger than $6~mm$ (corresponds to the flexural strain of $0.045$).

These two types of flexural experiments can together suggest that the PLA-CFPLA bi-layer composite combines the high stiffness and light-weight of CFPLA and the high flexibility and high strength of PLA. The combination of advantages of these two materials makes PLA-CFPLA a better choice for our structure and sample design.

\section{Additional point-pair measurement result}
\begin{figure}[!ht]
    \centering{
    \includegraphics[width=\linewidth]{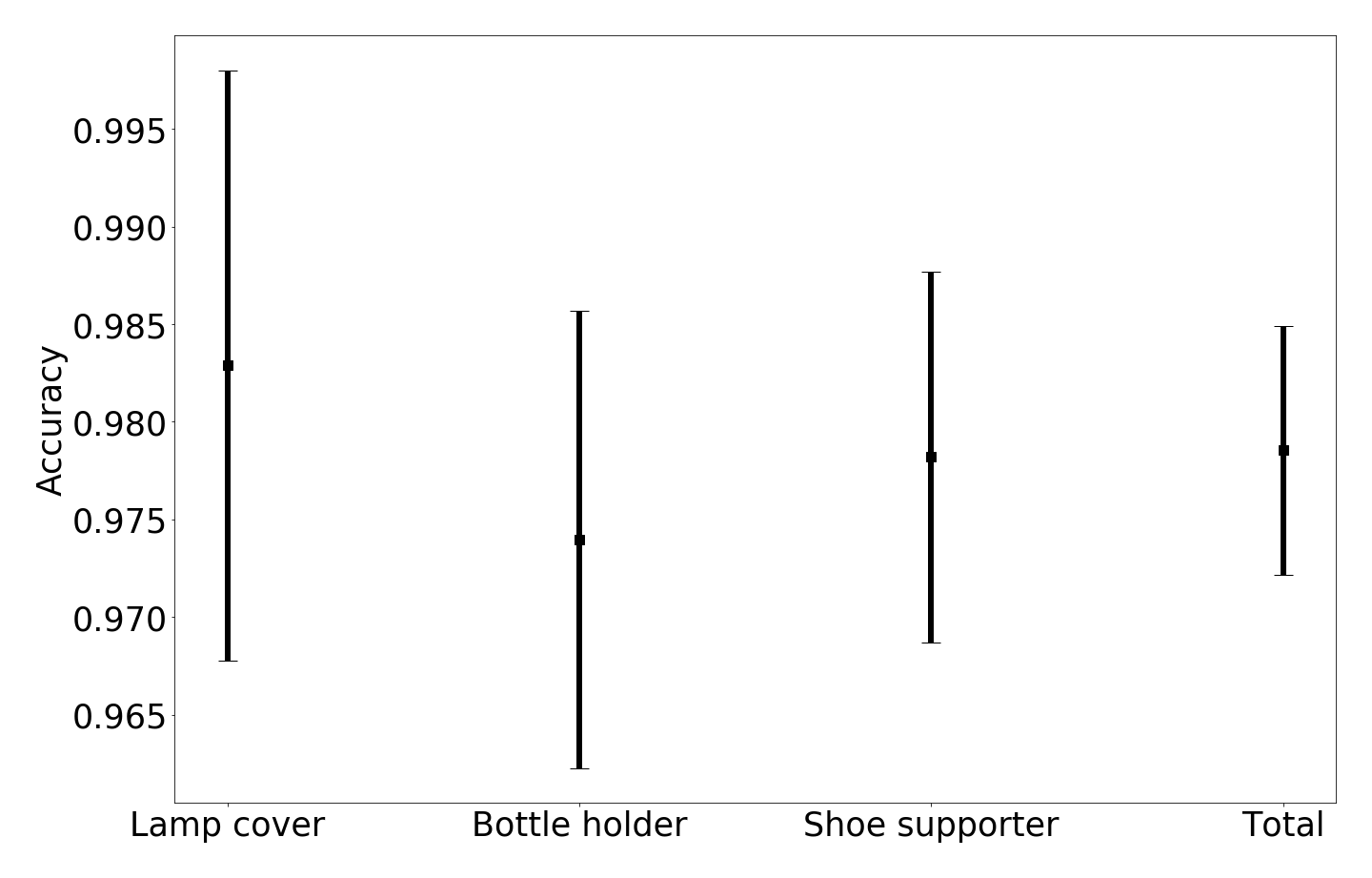}
    \caption{\yyred{The $95\%$ confidence interval of point-pair measurement results. }}
    \label{fig:point_pair}}
\end{figure}

We randomly picked several more point pairs on the designed applications and measured the distance difference between experiment and simulation (Table~\ref{tab:point_pair_measurement}). Figure~\ref{fig:point_pair} shows \yyred{the $95\%$ confidence interval from sample data for each designed application. The abstract indicates the $95\%$ confidence interval for the accuracy characterization based on all sample data for the designed applications.}

\clearpage

\begin{small}
\begin{table*}[ht!]
\vspace{-57.5em}
    \centering{
    \caption{Additional simulation-experiment error data for the designed applications}
    \vspace{-2mm}
    \begin{tabular}{c|c|c|c|c|c|c|c|c}
        \hline
        \multicolumn{3}{c|}{\tabincell{c}{Lamp Cover}} & \multicolumn{3}{c|}{\tabincell{c}{Bottle Holder}} & \multicolumn{3}{c}{\tabincell{c}{Shoe Supporter}} \\
        \hline
        \tabincell{c}{Experiment \\ (mm)} & \tabincell{c}{Simulation \\ (mm)} & \tabincell{c}{Error (\%)} & \tabincell{c}{Experiment \\ (mm)} & \tabincell{c}{Simulation \\ (mm)} & \tabincell{c}{Error (\%)} & \tabincell{c}{Experiment \\ (mm)} & \tabincell{c}{Simulation \\ (mm)} & \tabincell{c}{Error (\%)} \\
        \hline
        103     &102.8  &       0.19 & 71      &68.4   &       3.66 & 113     &110.78 &1.96\\
        102     &101.49 &       0.50 & 161     &157.27 &       2.32 & 115     &110.78 &3.67\\
        82      &81.34  &       0.80 & 68      &65     &       4.41 & 154     &155.32 &0.86\\
        165     &166.2  &       0.73 & 160     &155.34 &    2.91 & 157     &157.41 &0.26\\
        105.5   &104.54 &       0.91 & 159     &155.34 &       2.30 & - & - & - \\
        \hline
    \end{tabular}
    \label{tab:point_pair_measurement}}
\end{table*}
\end{small}

\clearpage

%% file: MaterialCharacterization_FEA_4DPrinting.bbl
\begin{thebibliography}{40}
\expandafter\ifx\csname natexlab\endcsname\relax\def\natexlab#1{#1}\fi
\providecommand{\url}[1]{\texttt{#1}}
\providecommand{\href}[2]{#2}
\providecommand{\path}[1]{#1}
\providecommand{\DOIprefix}{DOI:}
\providecommand{\ArXivprefix}{arXiv:}
\providecommand{\URLprefix}{URL: }
\providecommand{\Pubmedprefix}{pmid:}
\providecommand{\doi}[1]{\href{http://dx.doi.org/#1}{\path{#1}}}
\providecommand{\Pubmed}[1]{\href{pmid:#1}{\path{#1}}}
\providecommand{\bibinfo}[2]{#2}
\ifx\xfnm\relax \def\xfnm[#1]{\unskip,\space#1}\fi
\bibitem[{Ngo et~al.(2018)Ngo, Kashani, Imbalzano, Nguyen, and
  Hui}]{ngo2018additive}
\bibinfo{author}{T.~Ngo}, \bibinfo{author}{A.~Kashani},
  \bibinfo{author}{G.~Imbalzano}, \bibinfo{author}{K.~Nguyen},
  \bibinfo{author}{D.~Hui},
\newblock \bibinfo{title}{Additive manufacturing (3{D} printing): {A} review of
  materials, methods, applications and challenges},
\newblock \bibinfo{journal}{Composites Part B: Engineering}
  \bibinfo{volume}{143} (\bibinfo{year}{2018}) \bibinfo{pages}{172--196}.
\bibitem[{Gao et~al.(2015)Gao, Zhang, Ramanujan, Ramani, Chen, Williams, Wang,
  Shin, Zhang, and Zavattieri}]{gao2015status}
\bibinfo{author}{W.~Gao}, \bibinfo{author}{Y.~Zhang},
  \bibinfo{author}{D.~Ramanujan}, \bibinfo{author}{K.~Ramani},
  \bibinfo{author}{Y.~Chen}, \bibinfo{author}{C.~B. Williams},
  \bibinfo{author}{C.~C.~L. Wang}, \bibinfo{author}{Y.~C. Shin},
  \bibinfo{author}{S.~Zhang}, \bibinfo{author}{P.~D. Zavattieri},
\newblock \bibinfo{title}{The status, challenges, and future of additive
  manufacturing in engineering},
\newblock \bibinfo{journal}{Computer-Aided Design} \bibinfo{volume}{69}
  (\bibinfo{year}{2015}) \bibinfo{pages}{65--89}.
\bibitem[{Hu et~al.(2019)Hu, Duan, Xing, Xu, Du, Zhou, Chen, and
  Shan}]{hu2019improved}
\bibinfo{author}{B.~Hu}, \bibinfo{author}{X.~Duan}, \bibinfo{author}{Z.~Xing},
  \bibinfo{author}{Z.~Xu}, \bibinfo{author}{C.~Du}, \bibinfo{author}{H.~Zhou},
  \bibinfo{author}{R.~Chen}, \bibinfo{author}{B.~Shan},
\newblock \bibinfo{title}{Improved design of fused deposition modeling
  equipment for 3{D} printing of high-performance peek parts},
\newblock \bibinfo{journal}{Mechanics of Materials} \bibinfo{volume}{137}
  (\bibinfo{year}{2019}) \bibinfo{pages}{103139}.
\bibitem[{Singh et~al.(2019)Singh, Singh, Singh, and
  Kumar}]{singh2019investigations}
\bibinfo{author}{R.~Singh}, \bibinfo{author}{G.~Singh},
  \bibinfo{author}{J.~Singh}, \bibinfo{author}{R.~Kumar},
\newblock \bibinfo{title}{Investigations for tensile, compressive and
  morphological properties of 3{D} printed functional prototypes of
  {PLA-PEKK-HAp-CS}},
\newblock \bibinfo{journal}{Journal of Thermoplastic Composite Materials}
  (\bibinfo{year}{2019}). \DOIprefix\doi{10.1177/0892705719870595}.
\bibitem[{Roskies et~al.(2017)Roskies, Fang, Abdallah, Charbonneau, Cohen,
  Jordan, Hier, Mlynarek, Tamimi, and Tran}]{roskies2017three}
\bibinfo{author}{M.~G. Roskies}, \bibinfo{author}{D.~Fang},
  \bibinfo{author}{M.-N. Abdallah}, \bibinfo{author}{A.~M. Charbonneau},
  \bibinfo{author}{N.~Cohen}, \bibinfo{author}{J.~O. Jordan},
  \bibinfo{author}{M.~P. Hier}, \bibinfo{author}{A.~Mlynarek},
  \bibinfo{author}{F.~Tamimi}, \bibinfo{author}{S.~D. Tran},
\newblock \bibinfo{title}{Three-dimensionally printed polyetherketoneketone
  scaffolds with mesenchymal stem cells for the reconstruction of
  critical-sized mandibular defects},
\newblock \bibinfo{journal}{The Laryngoscope} \bibinfo{volume}{127}
  (\bibinfo{year}{2017}) \bibinfo{pages}{E392--E398}.
\bibitem[{Tian et~al.(2016)Tian, Liu, Yang, Wang, and Li}]{tian2016interface}
\bibinfo{author}{X.~Tian}, \bibinfo{author}{T.~Liu}, \bibinfo{author}{C.~Yang},
  \bibinfo{author}{Q.~Wang}, \bibinfo{author}{D.~Li},
\newblock \bibinfo{title}{Interface and performance of 3{D} printed continuous
  carbon fiber reinforced pla composites},
\newblock \bibinfo{journal}{Composites Part A: Applied Science and
  Manufacturing} \bibinfo{volume}{88} (\bibinfo{year}{2016})
  \bibinfo{pages}{198--205}.
\bibitem[{Tibbits(2014)}]{tibbits20144d}
\bibinfo{author}{S.~Tibbits},
\newblock \bibinfo{title}{{4D printing: multi-material shape change}},
\newblock \bibinfo{journal}{Architectural Design} \bibinfo{volume}{84}
  (\bibinfo{year}{2014}) \bibinfo{pages}{116--121}.
\bibitem[{Ge et~al.(2013)Ge, Qi, and Dunn}]{ge2013active}
\bibinfo{author}{Q.~Ge}, \bibinfo{author}{H.~J. Qi}, \bibinfo{author}{M.~L.
  Dunn},
\newblock \bibinfo{title}{Active materials by four-dimension printing},
\newblock \bibinfo{journal}{Applied Physics Letters} \bibinfo{volume}{103}
  (\bibinfo{year}{2013}) \bibinfo{pages}{131901}.
\bibitem[{Kwok et~al.(2015)Kwok, Wang, Deng, Zhang, and Chen}]{kwok2015four}
\bibinfo{author}{T.~H. Kwok}, \bibinfo{author}{C.~C.~L. Wang},
  \bibinfo{author}{D.~Deng}, \bibinfo{author}{Y.~Zhang},
  \bibinfo{author}{Y.~Chen},
\newblock \bibinfo{title}{Four-dimensional printing for freeform surfaces:
  design optimization of origami and kirigami structures},
\newblock \bibinfo{journal}{Journal of Mechanical Design} \bibinfo{volume}{137}
  (\bibinfo{year}{2015}) \bibinfo{pages}{111413}.
\bibitem[{Deng and Chen(2015)}]{deng2015origami}
\bibinfo{author}{D.~Deng}, \bibinfo{author}{Y.~Chen},
\newblock \bibinfo{title}{Origami-based self-folding structure design and
  fabrication using projection based stereolithography},
\newblock \bibinfo{journal}{Journal of Mechanical Design} \bibinfo{volume}{137}
  (\bibinfo{year}{2015}) \bibinfo{pages}{021701}.
\bibitem[{An et~al.(2014)An, Miyashita, Tolley, Aukes, Meeker, Demaine,
  Demaine, Wood, and Rus}]{an2014end}
\bibinfo{author}{B.~An}, \bibinfo{author}{S.~Miyashita}, \bibinfo{author}{M.~T.
  Tolley}, \bibinfo{author}{D.~M. Aukes}, \bibinfo{author}{L.~Meeker},
  \bibinfo{author}{E.~D. Demaine}, \bibinfo{author}{M.~L. Demaine},
  \bibinfo{author}{R.~J. Wood}, \bibinfo{author}{D.~Rus},
\newblock \bibinfo{title}{{An end-to-end approach to making self-folded 3D
  surface shapes by uniform heating}},
\newblock in: \bibinfo{booktitle}{IEEE International Conference on Robotics and
  Automation}, \bibinfo{year}{2014}, pp. \bibinfo{pages}{1466--1473}.
\bibitem[{Leung et~al.(2019)Leung, Kwok, Li, Yang, Wang, and
  Chen}]{leung2019challenges}
\bibinfo{author}{Y.~S. Leung}, \bibinfo{author}{T.~H. Kwok},
  \bibinfo{author}{X.~Li}, \bibinfo{author}{Y.~Yang}, \bibinfo{author}{C.~C.~L.
  Wang}, \bibinfo{author}{Y.~Chen},
\newblock \bibinfo{title}{Challenges and status on design and computation for
  emerging additive manufacturing technologies},
\newblock \bibinfo{journal}{Journal of Computing and Information Science in
  Engineering} \bibinfo{volume}{19} (\bibinfo{year}{2019})
  \bibinfo{pages}{021013}.
\bibitem[{Davis et~al.(2016)Davis, Chen, Dickey, and Genzer}]{davis2016self}
\bibinfo{author}{D.~Davis}, \bibinfo{author}{B.~Chen}, \bibinfo{author}{M.~D.
  Dickey}, \bibinfo{author}{J.~Genzer},
\newblock \bibinfo{title}{Self-folding of thick polymer sheets using gradients
  of heat},
\newblock \bibinfo{journal}{Journal of Mechanisms and Robotics}
  \bibinfo{volume}{8} (\bibinfo{year}{2016}) \bibinfo{pages}{031014}.
\bibitem[{Deng et~al.(2017)Deng, Yang, Chen, Lan, and
  Tice}]{deng2017accurately}
\bibinfo{author}{D.~Deng}, \bibinfo{author}{Y.~Yang},
  \bibinfo{author}{Y.~Chen}, \bibinfo{author}{X.~Lan},
  \bibinfo{author}{J.~Tice},
\newblock \bibinfo{title}{Accurately controlled sequential self-folding
  structures by polystyrene film},
\newblock \bibinfo{journal}{Smart Materials and Structures}
  \bibinfo{volume}{26} (\bibinfo{year}{2017}) \bibinfo{pages}{085040}.
\bibitem[{Kwok and Chen(2017)}]{kwok2017gdfe}
\bibinfo{author}{T.~H. Kwok}, \bibinfo{author}{Y.~Chen},
\newblock \bibinfo{title}{{GDFE}: Geometry-driven finite element for
  four-dimensional printing},
\newblock \bibinfo{journal}{Journal of Manufacturing Science and Engineering}
  \bibinfo{volume}{139} (\bibinfo{year}{2017}) \bibinfo{pages}{111006}.
\bibitem[{Bakarich et~al.(2015)Bakarich, Gorkin, Panhuis, and
  Spinks}]{bakarich20154d}
\bibinfo{author}{S.~E. Bakarich}, \bibinfo{author}{R.~Gorkin},
  \bibinfo{author}{M.~H. Panhuis}, \bibinfo{author}{G.~M. Spinks},
\newblock \bibinfo{title}{{4D printing with mechanically robust, thermally
  actuating hydrogels}},
\newblock \bibinfo{journal}{Macromolecular Rapid Communications}
  \bibinfo{volume}{36} (\bibinfo{year}{2015}) \bibinfo{pages}{1211--1217}.
\bibitem[{Grijpma and Pennings(1994)}]{grijpma1994co}
\bibinfo{author}{D.~W. Grijpma}, \bibinfo{author}{A.~J. Pennings},
\newblock \bibinfo{title}{{(Co) polymers of L-lactide, 2. mechanical
  properties}},
\newblock \bibinfo{journal}{Macromolecular Chemistry and Physics}
  \bibinfo{volume}{195} (\bibinfo{year}{1994}) \bibinfo{pages}{1649--1663}.
\bibitem[{Maharana et~al.(2009)Maharana, Mohanty, and Negi}]{maharana2009melt}
\bibinfo{author}{T.~Maharana}, \bibinfo{author}{B.~Mohanty},
  \bibinfo{author}{Y.~S. Negi},
\newblock \bibinfo{title}{Melt--solid polycondensation of lactic acid and its
  biodegradability},
\newblock \bibinfo{journal}{Progress in Polymer Science} \bibinfo{volume}{34}
  (\bibinfo{year}{2009}) \bibinfo{pages}{99--124}.
\bibitem[{R.~Langer(1998)}]{langer1998drug}
\bibinfo{author}{R.~R.~Langer},
\newblock \bibinfo{title}{Drug delivery and targeting},
\newblock \bibinfo{journal}{Nature} \bibinfo{volume}{392}
  (\bibinfo{year}{1998}) \bibinfo{pages}{5--10}.
\bibitem[{Soares et~al.(2008)Soares, Moore, and
  Rajagopal}]{soares2008constitutive}
\bibinfo{author}{J.~S. Soares}, \bibinfo{author}{J.~E. Moore},
  \bibinfo{author}{K.~R. Rajagopal},
\newblock \bibinfo{title}{Constitutive framework for biodegradable polymers
  with applications to biodegradable stents},
\newblock \bibinfo{journal}{{ASAIO} Journal} \bibinfo{volume}{54}
  (\bibinfo{year}{2008}) \bibinfo{pages}{295--301}.
\bibitem[{Hayman et~al.(2014)Hayman, Bergerson, Miller, Moreno, and
  Moore}]{hayman2014effect}
\bibinfo{author}{D.~Hayman}, \bibinfo{author}{C.~Bergerson},
  \bibinfo{author}{S.~Miller}, \bibinfo{author}{M.~Moreno},
  \bibinfo{author}{J.~E. Moore},
\newblock \bibinfo{title}{{The effect of static and dynamic loading on
  degradation of PLLA stent fibers}},
\newblock \bibinfo{journal}{Journal of Biomechanical Engineering}
  \bibinfo{volume}{136} (\bibinfo{year}{2014}) \bibinfo{pages}{081006}.
\bibitem[{Khan and El-Sayed(2013)}]{khan2013phenomenological}
\bibinfo{author}{K.~A. Khan}, \bibinfo{author}{T.~El-Sayed},
\newblock \bibinfo{title}{A phenomenological constitutive model for the
  nonlinear viscoelastic responses of biodegradable polymers},
\newblock \bibinfo{journal}{Acta Mechanica} \bibinfo{volume}{224}
  (\bibinfo{year}{2013}) \bibinfo{pages}{287--305}.
\bibitem[{S{\"o}ntjens et~al.(2012)S{\"o}ntjens, Engels, Smit, and
  Govaert}]{sontjens2012time}
\bibinfo{author}{S.~H.~M. S{\"o}ntjens}, \bibinfo{author}{T.~A.~P. Engels},
  \bibinfo{author}{T.~H. Smit}, \bibinfo{author}{L.~E. Govaert},
\newblock \bibinfo{title}{{Time-dependent failure of amorphous poly-D,
  L-lactide: {I}nfluence of molecular weight}},
\newblock \bibinfo{journal}{Journal of the Mechanical Behavior of Biomedical
  Materials} \bibinfo{volume}{13} (\bibinfo{year}{2012})
  \bibinfo{pages}{69--77}.
\bibitem[{Eswaran et~al.(2011)Eswaran, Kelley, Bergstrom, and
  Giddings}]{eswaran2011material}
\bibinfo{author}{S.~K. Eswaran}, \bibinfo{author}{J.~A. Kelley},
  \bibinfo{author}{J.~S. Bergstrom}, \bibinfo{author}{V.~L. Giddings},
\newblock \bibinfo{title}{Material modeling of polylactide},
\newblock in: \bibinfo{booktitle}{SIMULIA Customer Conference, Barcelona,
  Spain, May}, \bibinfo{year}{2011}, pp. \bibinfo{pages}{17--19}.
\bibitem[{Bodaghi et~al.(2019)Bodaghi, Noroozi, Zolfagharian, Fotouhi, and
  Norouzi}]{bodaghi20194d}
\bibinfo{author}{M.~Bodaghi}, \bibinfo{author}{R.~Noroozi},
  \bibinfo{author}{A.~Zolfagharian}, \bibinfo{author}{M.~Fotouhi},
  \bibinfo{author}{S.~Norouzi},
\newblock \bibinfo{title}{{4D printing self-morphing structures}},
\newblock \bibinfo{journal}{Materials} \bibinfo{volume}{12}
  (\bibinfo{year}{2019}) \bibinfo{pages}{1353}.
\bibitem[{Diani et~al.(2009)Diani, Fayolle, and Gilormini}]{diani2009review}
\bibinfo{author}{J.~Diani}, \bibinfo{author}{B.~Fayolle},
  \bibinfo{author}{P.~Gilormini},
\newblock \bibinfo{title}{A review on the mullins effect},
\newblock \bibinfo{journal}{European Polymer Journal} \bibinfo{volume}{45}
  (\bibinfo{year}{2009}) \bibinfo{pages}{601--612}.
\bibitem[{Ionov(2011)}]{ionov2011soft}
\bibinfo{author}{L.~Ionov},
\newblock \bibinfo{title}{Soft microorigami: {S}elf-folding polymer films},
\newblock \bibinfo{journal}{Soft Matter} \bibinfo{volume}{7}
  (\bibinfo{year}{2011}) \bibinfo{pages}{6786--6791}.
\bibitem[{Shim et~al.(2012)Shim, Kim, Heo, Jeon, and Yang}]{shim2012controlled}
\bibinfo{author}{T.~S. Shim}, \bibinfo{author}{S.~H. Kim},
  \bibinfo{author}{C.~J. Heo}, \bibinfo{author}{H.~C. Jeon},
  \bibinfo{author}{S.~M. Yang},
\newblock \bibinfo{title}{Controlled origami folding of hydrogel bilayers with
  sustained reversibility for robust microcarriers},
\newblock \bibinfo{journal}{Angewandte Chemie International Edition}
  \bibinfo{volume}{51} (\bibinfo{year}{2012}) \bibinfo{pages}{1420--1423}.
\bibitem[{Stoychev et~al.(2013)Stoychev, Turcaud, Dunlop, and
  Ionov}]{stoychev2013hierarchical}
\bibinfo{author}{G.~Stoychev}, \bibinfo{author}{S.~Turcaud},
  \bibinfo{author}{J.~Dunlop}, \bibinfo{author}{L.~Ionov},
\newblock \bibinfo{title}{Hierarchical multi-step folding of polymer bilayers},
\newblock \bibinfo{journal}{Advanced Functional Materials} \bibinfo{volume}{23}
  (\bibinfo{year}{2013}) \bibinfo{pages}{2295--2300}.
\bibitem[{Jones and Ashby(2005)}]{jones2005engineering}
\bibinfo{author}{D.~R.~H. Jones}, \bibinfo{author}{M.~F. Ashby},
  \bibinfo{title}{{Engineering materials 2: {A}n introduction to
  microstructures, processing and design}}, \bibinfo{publisher}{Elsevier},
  \bibinfo{year}{2005}.
\bibitem[{Osborne(1969)}]{osborne1969shooting}
\bibinfo{author}{M.~R. Osborne},
\newblock \bibinfo{title}{On shooting methods for boundary value problems},
\newblock \bibinfo{journal}{Journal of Mathematical Analysis and Applications}
  \bibinfo{volume}{27} (\bibinfo{year}{1969}) \bibinfo{pages}{417--433}.
\bibitem[{Eilers and Marx(1996)}]{eilers1996flexible}
\bibinfo{author}{P.~H.~C. Eilers}, \bibinfo{author}{B.~D. Marx},
\newblock \bibinfo{title}{Flexible smoothing with {B}-splines and penalties},
\newblock \bibinfo{journal}{Statistical Science} \bibinfo{volume}{11}
  (\bibinfo{year}{1996}) \bibinfo{pages}{89--102}.
\bibitem[{Yu et~al.(2019)Yu, Zhang, Takizawa, Tezduyar, and
  Sasaki}]{yu2019anatomically}
\bibinfo{author}{Y.~Yu}, \bibinfo{author}{Y.~J. Zhang},
  \bibinfo{author}{K.~Takizawa}, \bibinfo{author}{T.~E. Tezduyar},
  \bibinfo{author}{T.~Sasaki},
\newblock \bibinfo{title}{Anatomically realistic lumen motion representation in
  patient-specific space--time isogeometric flow analysis of coronary arteries
  with time-dependent medical-image data},
\newblock \bibinfo{journal}{Computational Mechanics}  (\bibinfo{year}{2019}).
  \DOIprefix\doi{10.1007/s00466-019-01774-4}.
\bibitem[{Lubliner(2008)}]{lubliner2008plasticity}
\bibinfo{author}{J.~Lubliner}, \bibinfo{title}{Plasticity theory},
  \bibinfo{publisher}{Courier Corporation}, \bibinfo{year}{2008}.
\bibitem[{Syst{\'e}mes(2019)}]{systemes2019abaqus}
\bibinfo{author}{D.~Syst{\'e}mes},
\newblock \bibinfo{title}{{SIMULIA user assistance}},
\newblock \bibinfo{journal}{Dassault Syst{\`e}mes}  (\bibinfo{year}{2019}).
\bibitem[{Marlow(2003)}]{marlow2003general}
\bibinfo{author}{R.~S. Marlow},
\newblock \bibinfo{title}{A general first-invariant hyperelastic constitutive
  model},
\newblock \bibinfo{journal}{Constitutive Models for Rubber}
  (\bibinfo{year}{2003}) \bibinfo{pages}{157--160}.
\bibitem[{Ogden and Roxburgh(1999)}]{ogden1999pseudo}
\bibinfo{author}{R.~W. Ogden}, \bibinfo{author}{D.~G. Roxburgh},
\newblock \bibinfo{title}{A pseudo--elastic model for the {M}ullins effect in
  filled rubber},
\newblock \bibinfo{journal}{Proceedings of the Royal Society of London. Series
  A: Mathematical, Physical and Engineering Sciences} \bibinfo{volume}{455}
  (\bibinfo{year}{1999}) \bibinfo{pages}{2861--2877}.
\bibitem[{Govindarajan et~al.(2008)Govindarajan, Hurtado, and
  Mars}]{govindarajan2008simulation}
\bibinfo{author}{S.~M. Govindarajan}, \bibinfo{author}{J.~A. Hurtado},
  \bibinfo{author}{W.~V. Mars},
\newblock \bibinfo{title}{Simulation of {M}ullins effect and permanent set in
  filled elastomers using multiplicative decomposition},
\newblock in: \bibinfo{booktitle}{Constitutive Models for Rubber -
  Proceedings}, volume~\bibinfo{volume}{5}, \bibinfo{organization}{Balkema},
  \bibinfo{year}{2008}, p. \bibinfo{pages}{249}.
\bibitem[{Yeoh(1993)}]{yeoh1993some}
\bibinfo{author}{O.~H. Yeoh},
\newblock \bibinfo{title}{Some forms of the strain energy function for rubber},
\newblock \bibinfo{journal}{Rubber Chemistry and technology}
  \bibinfo{volume}{66} (\bibinfo{year}{1993}) \bibinfo{pages}{754--771}.
\bibitem[{Zhou et~al.(2018)Zhou, Cui, Zhang, Yang, Liu, and
  Sun}]{zhou2018graph}
\bibinfo{author}{J.~Zhou}, \bibinfo{author}{G.~Cui},
  \bibinfo{author}{Z.~Zhang}, \bibinfo{author}{C.~Yang},
  \bibinfo{author}{Z.~Liu}, \bibinfo{author}{M.~Sun},
\newblock \bibinfo{title}{Graph neural networks: {A} review of methods and
  applications},
\newblock \bibinfo{journal}{arXiv:1812.08434}  (\bibinfo{year}{2018}).

\end{thebibliography}
